\documentclass[dvips,12pt,a4paper]{article}
\usepackage{a4wide}
\usepackage{epsfig}
\usepackage{amsmath}
\usepackage{latexsym}

  \newlength{\absize}
\newcommand{\dd}{\mbox{{\rm d}}}

\newcommand{\thW}{\theta_{{\rm W}}}

\newcommand{\Lumint}{{\cal L}_{\rm int}}

\allowdisplaybreaks 

\catcode`@=11
\def\citer{\@ifnextchar [{\@tempswatrue\@citexr}{\@tempswafalse\@citexr[]}}
 
\def\@citexr[#1]#2{\if@filesw\immediate\write\@auxout{\string\citation{#2}}\fi
  \def\@citea{}\@cite{\@for\@citeb:=#2\do
    {\@citea\def\@citea{--\penalty\@m}\@ifundefined
       {b@\@citeb}{{\bf ?}\@warning
       {Citation `\@citeb' on page \thepage \space undefined}}%
\hbox{\csname b@\@citeb\endcsname}}}{#1}}
\catcode`@=12

\begin{document}
  \thispagestyle{empty}
  \pagestyle{empty}
  \renewcommand{\thefootnote}{\fnsymbol{footnote}}
\newpage\normalsize
    \pagestyle{plain}
    \setlength{\baselineskip}{4ex}\par
    \setcounter{footnote}{0}
    \renewcommand{\thefootnote}{\arabic{footnote}}
\newcommand{\preprint}[1]{%
  \begin{flushright}
    \setlength{\baselineskip}{3ex} #1
  \end{flushright}}
\renewcommand{\title}[1]{%
  \begin{center}
    \LARGE #1
  \end{center}\par}
\renewcommand{\author}[1]{%
  \vspace{2ex}
  {\Large
   \begin{center}
     \setlength{\baselineskip}{3ex} #1 \par
   \end{center}}}
\renewcommand{\thanks}[1]{\footnote{#1}}
\renewcommand{\abstract}[1]{%
  \vspace{2ex}
  \normalsize
  \begin{center}
    \centerline{\bf Abstract}\par
    \vspace{2ex}
    \parbox{\absize}{#1\setlength{\baselineskip}{2.5ex}\par}
  \end{center}}

\begin{flushright}
{\setlength{\baselineskip}{2ex}\par
{\tt University of Trieste, Dept.\ of Theoretical Physics} \\[1mm]
{\tt UTS-DFT-00-01}    \\[1mm]
{hep-ph/0003253} \\[2mm]
{March 2000}           \\
} 
\end{flushright}
\vspace*{4mm}
\vfill
\title{Contact interaction probes at the Linear Collider with 
polarized electron and positron beams\thanks{Partially supported 
by the Research Council of Norway,
and by MURST (Italian Ministry of University, 
Scientific Research and Technology).}}
\vfill
\author{
A.A. Babich$^{a}$,
P. Osland$^{b}$,
A.A. Pankov$^{a,c}$ {\rm and}
N. Paver$^{c}$}
\begin{center}
$^a$ Pavel Sukhoi Technical University, 
     Gomel, 246746 Belarus \\
$^b$ Department of Physics, University of Bergen, \\
     All\'{e}gaten 55, N-5007 Bergen, Norway \\
$^c$ Dipartimento di Fisica Teorica, Universit\`a di Trieste and \\
Istituto Nazionale di Fisica Nucleare, Sezione di Trieste,
Trieste, Italy
\end{center}
\vfill
\abstract
{For contact-interaction searches at the Linear Collider,
we discuss the advantages of polarizing both the electron and the positron
beams as compared with polarizing only the electron beam.
In particular, for the processes $e^+e^-\to \mu^+\mu^-$, $\tau^+\tau^-$,
$b\bar{b}$ and $c\bar{c}$ at a future $e^+e^-$ collider with $\sqrt{s}=0.5$
TeV we derive model-independent bounds on the four-fermion contact
interaction parameters from studies of the helicity cross sections.
}
\vspace*{20mm}
\setcounter{footnote}{0}
\vfill

\newpage
    \setcounter{footnote}{0}
    \renewcommand{\thefootnote}{\arabic{footnote}}
    \setcounter{page}{1}

\section{Introduction}

The possibility of constructing high energy polarized electron and positron 
beams is considered with great interest with regard to the physics 
programme at the Linear Collider (LC). Indeed, one of the most important 
advantages of initial beam polarization is that one can measure spin-dependent
observables, which represent the most direct probes of the fermion 
helicity dependence of the electroweak interactions. Consequently, one would
expect a substantial gain in the sensitivity to the features of possible
non-standard interactions and, in particular, stringent constraints on the
individual new coupling constants could be derived from the data analysis by 
looking for deviations of cross sections from the Standard Model (SM) 
predictions. 

Here, we will consider the process of fermion pair production 
($f\ne e$, $t$)
\begin{equation}
e^++e^-\to f+\bar{f} \label{proc}
\end{equation}
at a future Linear Collider with longitudinally polarized electron 
{\it and} positron beams, and discuss the sensitivity of the 
measurable helicity cross sections to the $SU(3)\times SU(2)\times U(1)$ 
symmetric $eeff$ contact-interaction Lagrangian with helicity-conserving 
and flavor-diagonal fermion currents \cite{Eichten}:       
\begin{equation}
{\cal L}=\sum_{\alpha\beta}\frac{g^2_{\rm eff}}{\Lambda^2_{\alpha\beta}}
\eta_{\alpha\beta}\left(\bar e_{\alpha}\gamma_\mu e_{\alpha}\right)
\left(\bar f_{\beta}\gamma^\mu f_{\beta}\right).
\label{lagra}
\end{equation}
In Eq.~(\ref{lagra}), generation and color indices have been suppressed,
$\alpha,\beta={\rm L,R}$ indicate left- or right-handed helicities, and the
parameters $\eta_{\alpha\beta}=\pm 1,0$ specify the chiral structure of the
individual interactions. Conventionally, one takes $g^2_{\rm eff}=4\pi$ as a 
reminder that the new interaction, originally proposed for compositeness, 
would become strong at $\sqrt s\sim\Lambda_{\alpha\beta}$. 
Actually, in a more general sense, 
${\cal L}$ should be considered as an effective Lagrangian which represents 
the leading, lowest dimensional, parameterization at the `low-energy' $E$ at
which we make measurements, of some non-standard interaction acting at a much
larger energy scale $\Lambda\gg E$. For example, in addition to the remnant 
compositeness binding force, this is the case of a variety of interactions 
generated by the exchange of very heavy 
objects with masses much larger than the Mandelstam variables of the
considered
process (\ref{proc}), such as the exchanges of a $Z^\prime$ with a few TeV 
mass \cite{barger} and of a heavy leptoquark \cite{altarelli}. In 
this effective framework, therefore, 
with the assumed conventional values of $\eta$'s and $g^2_{\rm eff}$, the 
scales $\Lambda_{\alpha\beta}$ in Eq.~(\ref{lagra}) define a standard 
to compare the reach of different new-physics searches in the process 
(\ref{proc}). 

Clearly, ${\cal L}$ should manifest itself by deviations of observables from 
the SM theoretical predictions. 
The sensitivity of measurements to the new coupling
constants, or, equivalently, the experimentally attainable reach in the free 
mass scales $\Lambda_{\alpha\beta}$, can be assessed by the numerical
comparison of such deviations to the expected experimental accuracies. 

For a given flavor $f$, Eq.~(\ref{lagra}) defines eight individual,
independent models corresponding to the combinations of the four chiralities
$\alpha,\beta$ with the $\pm$ signs of the $\eta$'s. 
However, in general, an observed
contact interaction could be any linear combination of these models, and this
leads to the complicated situation in which the aforementioned deviations of 
observables from the SM predictions simultaneously depend on all four-fermion 
effective couplings. A simplified, and commonly adopted, procedure is to
assume a non-zero value for only one parameter at a time and constrain it by
essentially a $\chi^2$ fit analysis, keeping the remaining parameters 
set equal to zero. In this way, tests of the individual models are obtained.

On the other hand, a general, model-independent, analysis must simultaneously 
include all terms of 
Eq.~(\ref{lagra}) as free parameters and, at the same time, must allow to 
disentangle their contributions to the basic observables so as to avoid 
potential cancellations between different contributions. 
Such cancellations can make the 
constraints considerably weaker or even spoil them. 
For this purpose, the longitudinal polarization of initial beams 
offers the possibility of experimentally 
separating from the data the individual helicity cross sections of process 
(\ref{proc}), each one being directly related to a single $eeff$ contact term 
and, therefore, depending on the minimal set of free independent parameters. 
The approach we adopt here uses as basic observables two particular, 
polarized, integrated cross sections that allow to
reconstruct the four helicity amplitudes {\it via} linear combinations of 
measurements at different beam polarizations.\footnote{Integrated observables
should be of advantage in the case of limited experimental statistics.} 
Moreover, in the definition of such integrated observables, optimal 
kinematical regions can be chosen to maximize the sensitivity to the 
individual four-fermion contact interactions. 

This kind of analysis, and the determination of the corresponding reach on 
$\Lambda_{\alpha\beta}$, was applied in 
Ref.~\cite{BOPP-99} for the LC with $\sqrt s= 0.5\ {\rm TeV}$ and only 
the electron beam longitudinally polarized, making standard
assumptions on the luminosity and on the expected systematic 
uncertainties on the cross section of 
process (\ref{proc}) for the different flavors. Indeed, longitudinal 
polarization of one beam is by itself already sufficient to disentangle the 
helicity cross sections from the data, if at least two values of
the polarization are available, {\it e.g.}, $\pm \vert P_e\vert$. In what 
follows, we extend the analysis of Ref.~\cite{BOPP-99} and discuss the case 
where also positron beam longitudinal polarization is available at the LC with
the same c.m.\ energy. 
Specifically, after giving the main definitions and briefly reviewing the 
procedure and findings for the sensitivity on $\Lambda_{\alpha\beta}$ 
obtained in \cite{BOPP-99}, we start by considering the effect of 
the uncertainty on the electron beam polarization that was disregarded there. 
We then consider the case of 
both electron and positron longitudinal polarizations, including in the 
analysis also the uncertainty on these polarizations.
\section{Separation of the helicity cross sections}

In Eq.~(\ref{proc}) we limit ourselves to the cases $f\neq e, t$ and make the  
approximation of negligible fermion mass with respect to the c.m.\ energy
$\sqrt s$. Then, the amplitude for $e^+e^-\to f\bar{f}$ is determined by 
the Born, 
$s$-channel, $\gamma$ and $Z$ exchanges plus the contact-interaction term of 
Eq.~(\ref{lagra}). With $P_e$ and $P_{\bar e}$ the longitudinal polarizations 
of the beams, and $\theta$ the angle between the incoming electron and the
outgoing fermion in the c.m.\ frame, the differential cross section 
reads \cite{Zeppenfeld2}: 
\begin{equation}
\frac{\dd\sigma}{\dd\cos\theta}
=\frac{3}{8}
\left[(1+\cos\theta)^2 {\sigma}_+
+(1-\cos\theta)^2 {\sigma}_-\right].
\label{cross}
\end{equation}
In terms of helicity cross sections $\sigma_{\alpha\beta}$ (with
$\alpha,\beta={\rm L,R}$): 
\begin{eqnarray}
{\sigma}_{+}&=&\frac{1}{4}\,
\left[(1-P_e)(1+P_{\bar{e}})\,\sigma_{\rm LL}
+(1+P_e)(1- P_{\bar{e}})\,\sigma_{\rm RR}\right]\nonumber \\
&=&\frac{D}{4}\,\left[(1-P_{\rm eff})\,\sigma_{\rm LL}
+(1+P_{\rm eff})\,\sigma_{\rm RR}\right], 
\label{s+} \\
{\sigma}_{-}&=&\frac{1}{4}\,
\left[(1-P_e)(1+ P_{\bar{e}})\,\sigma_{\rm LR}
+(1+P_e)(1-P_{\bar{e}})\,\sigma_{\rm RL}\right] \nonumber \\
&=&
\frac{D}{4}\,\left[(1-P_{\rm eff})\,\sigma_{\rm LR}
+(1+P_{\rm eff})\,\sigma_{\rm RL}\right], \label{s-}
\end{eqnarray}
where 
\begin{equation}
P_{\rm eff}=\frac{P_e-P_{\bar{e}}}{1-P_eP_{\bar{e}}} 
\label{pg}
\end{equation}
is the effective polarization \cite{Flottmann-Omori}, 
$\vert P_{\rm eff}\vert\leq 1$, and 
$D=1-P_eP_{\bar{e}}$. Obviously, for unpolarized positrons 
$P_{\rm eff}\rightarrow P_e$ and $D\rightarrow 1$.
It should be noted that with $P_{\bar{e}}\ne0$, $\vert P_{\rm eff}\vert$
can be larger than $|P_e|$.
Moreover, in Eqs.~(\ref{s+}) and (\ref{s-}):  
\begin{equation}
\sigma_{\alpha\beta}=N_C\sigma_{\rm pt}
\vert A_{\alpha\beta}\vert^2,
\label{helcross}
\end{equation}
where $N_C\approx 3(1+\alpha_s/\pi)$ for quarks and $N_C=1$ for leptons, 
respectively, and $\sigma_{\rm pt}\equiv\sigma(e^+e^-\to\gamma^\ast\to l^+l^-)
=(4\pi\alpha^2)/(3s)$.
The helicity amplitudes $A_{\alpha\beta}$ can be written as
\begin{equation}
A_{\alpha\beta}=Q_e Q_f+g_\alpha^e\,g_\beta^f\,\chi_Z+
\frac{s\eta_{\alpha\beta}}{\alpha\Lambda_{\alpha\beta}^2},
\label{amplit}
\end{equation}
where $\chi_Z=s/(s-M^2_Z+iM_Z\Gamma_Z)$ is the gauge boson propagator,
$g_{\rm L}^f=(I_{3L}^f-Q_f s_W^2)/s_W c_W$ and 
$g_{\rm R}^f=-Q_f s_W^2/s_W c_W$ 
are the SM left- and right-handed fermion couplings of the $Z$
with $s_W^2=1-c_W^2\equiv \sin^2\theta_W$ 
and $Q_f$ the fermion electric charge.

Our analysis focuses on the helicity cross sections that, as the above 
relations clearly show, directly relate to the individual contact 
interactions in Eq.~(\ref{lagra})
with definite chiralities and, accordingly, lead to a 
model-independent analysis where all terms in this equation are taken into
account as completely free parameters with no danger of accidental
compensations. To disentangle the various contributions in Eqs.~(\ref{s+}) and
(\ref{s-}), one simply has to make measurements at two different values of
the polarizations (a minimum of four measurements is needed). For example, two
convenient sets of values for the polarizations, that we will use in the 
sequel, would be $P_e=\pm P_1$ and $P_{\bar e}=\mp P_2$ ($P_{1,2}>0$) or, 
alternatively, $P_{\rm eff}=\pm P$ and $D$ fixed. 
The corresponding solutions of 
Eqs.~(\ref{s+}) and (\ref{s-}) read:
\begin{eqnarray}
\label{SLL}
\sigma_{\rm LL}
&=&\frac{1}{D}\left[
- \frac{1-P}{P}\sigma_{+}(P)
+\frac{1+P}{P}\sigma_{+}(-P) \right], \\
\label{SRR}
\sigma_{\rm RR}
&=&\frac{1}{D}\left[\frac{1+P}{P}\sigma_{+}(P) 
- \frac{1-P}{P}\sigma_{+}(-P)\right],
\end{eqnarray} 
with $\sigma_{\rm LR}$ and $\sigma_{\rm RL}$ obtained from 
$\sigma_{\rm LL}$ and $\sigma_{\rm RR}$, respectively, 
replacing $\sigma_{+}$ by $\sigma_{-}$.

Actually, for the purpose of optimizing the resulting 
bounds on $\Lambda_{\alpha\beta}$, one can more generally define the polarized
cross sections integrated over the {\it a priori} arbitrary kinematical ranges
($-1,\ z^*$) and ($z^*,\ 1$) \cite{BOPP-99}: 
\begin{eqnarray}
\label{sigma1}
\sigma_1(z^*, P, D)
&\equiv&\int_{z^*}^1\frac{\dd\sigma}{\dd\cos\theta}\dd\cos\theta
=\frac{1}{8}\left\{\left[8-(1+z^*)^3\right]\sigma_++(1-z^*)^3
\sigma_-\right\}, \\
\label{sigma2}
\sigma_2(z^*, P,{D})
&\equiv&\int^{z^*}_{-1}\frac{\dd\sigma}{\dd\cos\theta}\dd\cos\theta
=\frac{1}{8}\left\{(1+z^*)^3\sigma_++
\left[8-(1-z^*)^3\right]\sigma_-\right\}.  
\end{eqnarray}
For simplicity of notations, the polarization dependence of $\sigma_{\pm}$ on
the right-hand sides of Eqs.~(\ref{sigma1}) and (\ref{sigma2}) has been
suppressed. As abbreviations, we introduce
\begin{equation}
\label{Eq:a-b}
a(z^*)=\frac{8-(1-z^*)^3}{6(1-{z^*}^2)}, \qquad
b(z^*)=-\frac{(1-z^*)^3}{6(1-{z^*}^2)}.
\end{equation}
By solving Eqs.~(\ref{sigma1}) and (\ref{sigma2}) one obtains 
$\sigma_+$ and $\sigma_-$ from the measurement of $\sigma_1$ and $\sigma_2$: 
\begin{eqnarray}
\label{sigmap}
\sigma_+
&=&\left[a(z^*)\sigma_1(z^*,P,D) 
        +b(z^*)\sigma_2(z^*,P,D)\right], \\
\label{sigmam}
\sigma_-
&=&\left[b(-z^*)\sigma_1(z^*,P,D)
        +a(-z^*)\sigma_2(z^*,P,D)\right].
\end{eqnarray}
Thus, according to this procedure, $\sigma_{1,2}(z^*, P, D)$ play the 
role of a basic set of integrated polarized observables to be measured. As a 
second step, the corresponding cross sections $\sigma_{\pm}$ are constructed 
using the relations (\ref{sigmap}) and (\ref{sigmam}) and the experimental
values of the helicity cross sections $\sigma_{\alpha\beta}$ are finally 
determined from the linear system of equations (\ref{SLL})--(\ref{SRR}). 
Moreover, the value of $z^*$ is taken as an input parameter related to given 
experimental conditions, that can be tuned in order to get maximal sensitivity
of the helicity cross sections $\sigma_{\alpha\beta}$ to the mass scales 
$\Lambda_{\alpha\beta}$ we want to constrain.    

For comparison, we recall the conventional observables,
the total cross section $\sigma$ and the various asymmetries.
These are generally given, according to Eqs.~(\ref{cross})--(\ref{pg}), by 
\begin{equation}
\label{canon}
\sigma
={\sigma}_{+}+{\sigma}_{-}
=\frac{D}{4}\left[(1-P_{\rm eff})(\sigma_{\rm LL}+\sigma_{\rm LR})
+(1+P_{\rm eff})(\sigma_{\rm RR}+\sigma_{\rm RL})\right]; 
\end{equation}
and
\begin{eqnarray}
\sigma A_{\rm FB}
&\equiv&\sigma_{\rm F}-\sigma_{\rm B}
=\frac{3}{4}\left({\sigma}_{+}-{\sigma}_{-}\right) \nonumber \\
&=&\frac{3}{16}D\left[(1-P_{\rm eff})(\sigma_{\rm LL}-\sigma_{\rm LR})
+(1+P_{\rm eff})(\sigma_{\rm RR}-\sigma_{\rm RL})\right];
\end{eqnarray}
with 
\begin{equation}
\label{sfb}
\sigma_{\rm F}=\sigma_1(z^*=0)
=\int_{0}^{1}(\dd\sigma/\dd\cos\theta)\dd\cos\theta; \quad
\sigma_{\rm B}=\sigma_2(z^*=0)
=\int_{-1}^{0}(\dd\sigma/\dd\cos\theta)\dd\cos\theta, 
\end{equation}
and $P_{\rm eff}\to 0$, $D\to 1$ for unpolarized beams. 
For the case of polarized beams, one has also the left-right asymmetry 
\begin{equation}
A_{\rm LR}
=\frac{\sigma_{\rm L}-\sigma_{\rm R}}{\sigma_{\rm L}+\sigma_{\rm R}}
=\frac{(\sigma_{\rm LL}+\sigma_{\rm LR})-(\sigma_{\rm RL}+\sigma_{\rm RR})}
{\sigma_{\rm LL}+\sigma_{\rm LR}+\sigma_{\rm RL}+\sigma_{\rm RR}},  
\label{alr}
\end{equation}
and the combined left-right forward-backward asymmetry 
\begin{equation}
A_{\rm LR,FB}
=\frac{(\sigma_{\rm L}^{\rm F}-\sigma_{\rm R}^{\rm F})
-(\sigma_{\rm L}^{\rm B}-\sigma_{\rm R}^{\rm B})}
{(\sigma_{\rm L}^{\rm F}+\sigma_{\rm R}^{\rm F})+
(\sigma_{\rm L}^{\rm B}+\sigma_{\rm R}^{\rm B})}=
\frac{3}{4}\,
\frac{\sigma_{\rm LL}-\sigma_{\rm RR}+\sigma_{\rm RL}-\sigma_{\rm LR}}
{\sigma_{\rm LL}+\sigma_{\rm RR}+\sigma_{\rm RL}+\sigma_{\rm LR}},
\label{afbpol}
\end{equation}
where $\sigma_{\rm L}$ and $\sigma_{\rm R}$ denote the cross sections with
left-handed and right-handed electrons and unpolarized positrons.

In the numerical analysis, radiative corrections including initial- and
final-state radiation are taken into account by 
means of the program ZFITTER \cite{zfitter}, which has to be used along with 
ZEFIT, adapted to the present discussion, with $m_{\rm top}=175$~GeV and
$m_H=100$~GeV. One-loop SM electroweak corrections are accounted for by 
improved Born amplitudes \cite{Hollik,Altarelli2}, such that the form of the 
previous formulae remains the same. Concerning initial-state radiation, a cut 
on the energy of the emitted photon $\Delta=E_\gamma/E_{\rm beam}=0.9$ is 
applied for $\sqrt s=0.5\ {\rm TeV}$
in order to avoid the radiative return to the $Z$ peak, 
and increase the signal originating from 
the contact interaction \cite{Djouadi}.  

\section{Sensitivity of observables and their optimization}

Given the current bounds on $\Lambda_{\alpha\beta}$, 
of the order of several TeV \cite{Lambda}, 
at the LC c.m.\ energy $\sqrt{s}=0.5$ TeV the
characteristic suppression factor $s/\Lambda^2$ in Eq.~(\ref{amplit}) 
is such that we can only look at indirect
manifestations of the contact interaction (\ref{lagra}) as 
deviations from the SM predictions. In this case, we can assess the
sensitivity of process (\ref{proc}) to the couplings in (\ref{lagra}), that
determines the corresponding reach on $\Lambda_{\alpha\beta}$, 
on the basis of the foreseen
experimental accuracy on the helicity cross sections $\sigma_{\alpha\beta}$. 
As stressed previously, the knowledge of the latter allows a
model-independent analysis, where all the contact interactions are disentangled
and therefore can be taken into account as free parameters simultaneously.    

Specifically, we define the `significance' of each helicity cross
section by the ratio
\begin{equation}
\label{signif}
{\cal S}(\sigma_{\alpha\beta})
=\frac{|\Delta\sigma_{\alpha\beta}|}{\delta\sigma_{\alpha\beta}}, 
\end{equation}
where $\Delta\sigma_{\alpha\beta}$ are the deviations from the SM prediction 
due to (\ref{lagra}), dominated for $\sqrt s\ll \Lambda_{\alpha\beta}$ by the 
linear interference term 
\begin{equation}
\Delta\sigma_{\alpha\beta}\equiv
\sigma_{\alpha\beta}-\sigma_{\alpha\beta}^{\rm SM}\simeq
2 N_C\, \sigma_{\rm pt}
\left(Q_e\, Q_f+g_{\alpha}^e\, g_{\beta}^f\,\chi_Z\right)
\frac{s\eta_{\alpha\beta}}{\alpha\Lambda_{\alpha\beta}^2},
\label{deltasig}
\end{equation}
and $\delta\sigma_{\alpha\beta}$ denotes the expected experimental uncertainty
on $\sigma_{\alpha\beta}$, combining statistical and systematic uncertainties.

In the procedure of determining helicity amplitudes {\it via} the integrated
polarized cross sections $\sigma_{1,2}$ outlined in the previous
section 
(see Eqs.~(\ref{SLL}), (\ref{SRR}), (\ref{sigmap}) and (\ref{sigmam})), 
adding all uncertainties in quadrature and neglecting for the moment 
the systematic uncertainty on the electron and positron polarizations, 
one can write: 
\begin{eqnarray}
\label{uncet1}
\left(\delta \sigma_{\rm LL}\right)^2
&=&a^2(z^*)\left[
\left(\frac{1-P}{P D}\right)^2 
(\delta\sigma_1 (z^*,P))^2
+
\left(\frac{1+P}{P D}\right )^2 
(\delta\sigma_1 (z^*,-P))^2
\right]
\nonumber \\
&+&b^2(z^*)\left[ 
\left ( \frac{1-P}{P D} \right )^2 
(\delta\sigma_2 (z^*,P))^2
+
\left ( \frac{1+P}{P D} \right )^2 
(\delta\sigma_2 (z^*,-P))^2
\right], 
\end{eqnarray}
\begin{eqnarray}
\label{uncet2}
\left(\delta \sigma_{\rm LR}\right)^2
&=&b^2(-z^*)\left[
\left(\frac{1-P}{P D} \right )^2 
(\delta\sigma_1(z^*,P))^2
+
\left (\frac{1+P}{P D}\right )^2 
(\delta\sigma_1 (z^*,-P))^2 
\right]
\nonumber \\
&+&a^2(-z^*) \left[
\left(\frac{1-P}{P D} \right )^2 
(\delta\sigma_2(z^*,P))^2
+
\left(\frac{1+P}{P D}\right)^2 
(\delta\sigma_2(z^*,-P))^2
\right], 
\end{eqnarray}
where $a$ and $b$ are given by Eq.~(\ref{Eq:a-b}).
For simplicity of notations, the dependence of $\delta\sigma_{1,2}$ on $D$ has
not been explicitly indicated. One can derive explicit expressions for
$\delta\sigma_{\rm RR}$ and $\delta\sigma_{\rm RL}$ from 
$\delta\sigma_{\rm LL}$ and 
$\delta\sigma_{\rm LR}$, respectively, by the replacement in the 
above equations of
$\pm P\to \mp P$ in $\delta\sigma_i (z^*,\pm P))$ but not in the corresponding
prefactors. 

Combining in quadrature statistical and systematic uncertainties on 
$\sigma_{1,2}$, one finds:
\begin{equation}
\label{delsi1}
(\delta\sigma_i)^2\simeq(\delta\sigma_i^{\rm SM})^2
=\frac{\sigma_i^{\rm SM}}{\epsilon\, \Lumint}
+\left(\delta^{\rm sys}\sigma_i^{\rm SM}\right)^2, \qquad i=1,2.
\end{equation}
For our numerical analysis we shall assume the commonly used reference values 
of the identification efficiencies, $\epsilon$, and the 
systematic uncertainties, $\delta^{\rm sys}$, for the various fermionic 
channels \cite{Damerell}: $\epsilon=95\%$ and $\delta^{\rm sys}=0.5\%$ 
for $l^+l^-$; $\epsilon=60\%$ and $\delta^{\rm sys}=1\%$ for $b\bar{b}$; 
$\epsilon=35\%$ and $\delta^{\rm sys}=1.5\%$ for $c\bar{c}$. Notice that, as 
a simplification, we take the same $\delta^{\rm sys}$ for both $i=1$ and 2, 
and independent of $z^*$ in the relevant angular range. Concerning the
statistical uncertainty, we consider the LC with $\sqrt{s}=0.5$~TeV, 
$\Lumint=50\ \mbox{fb}^{-1}$ and $\Lumint=500\ \mbox{fb}^{-1}$
(half for each polarization orientation), and 
a fiducial experimental angular range $|\cos\theta|\le 0.99$.   

Finally, as regards optimization of the bounds on contact-interaction 
couplings, which corresponds to the maximum value of the `significance' 
defined in 
Eq.~(\ref{signif}), one may notice from the equations above that the
uncertainties $\delta\sigma_{\alpha\beta}$ depend on the, {\it a priori} free, 
kinematical parameter $z^*$ in the definition of the polarized cross sections 
$\sigma_i$. Conversely, by definition, the deviations from the SM 
$\Delta\sigma_{\alpha\beta}$ in Eq.~(\ref{deltasig}) are $z^*$-independent. 
Therefore, optimization can be achieved by choosing $z^*=z^*_{\rm opt}$ where 
$\delta\sigma_{\alpha\beta}$ becomes minimum, so that the corresponding 
sensitivity has a maximum and determines the highest bound on 
the corresponding mass scale $\Lambda_{\alpha\beta}$. 
The $z^*$ dependence of the statistical uncertainties 
$\delta\sigma^{\rm stat}_{\alpha\beta}$ in the right-hand side of 
(\ref{delsi1}) can be approximated by that corresponding to the 
known SM cross sections for the process (\ref{proc}) 
and the value of ${\cal L}_{\rm int}$. In the case of low 
luminosity where the statistical uncertainty dominates, 
this SM-determined $z^*$ behaviour 
can be used for a simple, first determination of $z^*_{\rm opt}$ 
for the various helicity amplitudes \cite{BOPP-99}.
In the general case where 
statistical and systematic uncertainties are comparable, the optimal $z^*$ 
must be determined by a more complex numerical analysis taking into account
the relevant experimental details.    

\section{Polarization uncertainty and two polarized beams}

In order to assess the effects on the $\delta\sigma_{\alpha\beta}$ 
due to the systematic uncertainties $\delta P_e$ and $\delta P_{\bar{e}}$ 
on the $e^-$ and $e^+$ polarizations respectively, we must supplement by
appropriate terms Eqs.~(\ref{uncet1}) and (\ref{uncet2}) and the similar ones 
for the remaining helicity amplitudes. 
From the formulae in Sec.~2, one can see that finite values of 
$\delta P_e$ and $\delta P_{\bar e}$ will influence the extraction 
of the helicity cross sections $\sigma_{\alpha\beta}$
through the prefactors of Eqs.~(\ref{SLL}), (\ref{SRR}),
(\ref{sigmap}) and (\ref{sigmam}), as well as through 
the dependence of $\sigma_{1,2}$ on $P$ and $D$. 
Clearly, a complete assessment of the latter effect would require 
detailed knowledge of the structure of the overall systematic uncertainty 
in terms of the different, individual sources, that is not available 
at present. 
For the sake of simplicity, we model the systematic uncertainty by assuming 
that such an effect can be considered as
already included in the systematic uncertainties 
$\delta\sigma_i^{\rm sys}$ introduced in Eq.~(\ref{delsi1}), 
regardless of the values of $\delta P_e$ and $\delta P_{\bar e}$
(and $P_e$ and $P_{\bar e}$) considered in our discussion.
Then, we treat $\sigma_{1,2}$, $P_e$ and $P_{\bar e}$ in 
Eqs.~(\ref{SLL}) and (\ref{SRR}) as if they were independent measurables,
and in this spirit, we combine the additional contribution 
to the uncertainty, $\delta\sigma_{\alpha\beta}^{\rm pol}$, 
again in quadrature with the $\delta\sigma_{\alpha\beta}$ determined 
from the expressions (\ref{uncet1}) and (\ref{uncet2}). 
Thus: 
\begin{equation}
(\delta \sigma_{\alpha \beta})^2 \Rightarrow
(\delta \sigma_{\alpha \beta})^2 
+ \left(\delta \sigma_{\alpha \beta}^{\rm pol}\right)^2.
\label{poldelta}
\end{equation}
Under the above assumptions, we obtain
\begin{eqnarray}
\label{Eq:uncert-P}
\left(\delta \sigma_{\rm LL}^{\rm pol}\right)^2
&=&[f(z^{\ast},P)(1+P_{\bar e}P^2)-f(z^{\ast},-P)(1-P_{\bar e}P^2)]^2
\left(\frac{\delta P_e}{D^2P^2}\right)^2 \nonumber \\
&+&[f(z^{\ast},P)(1-P_{e}P^2)-f(z^{\ast},-P)(1+P_{e}P^2)]^2
\left(\frac{\delta P_{\bar e}}{D^2P^2}\right)^2, \nonumber \\
\left(\delta \sigma_{\rm RR}^{\rm pol}\right)^2
&=&[f(z^{\ast},P)(1-P_{\bar e}P^2)-f(z^{\ast},-P)(1+P_{\bar e}P^2)]^2
\left(\frac{\delta P_e}{D^2P^2}\right)^2 \nonumber \\
&+&[f(z^{\ast},P)(1+P_{e}P^2)-f(z^{\ast},-P)(1-P_{e}P^2)]^2
\left(\frac{\delta P_{\bar e}}{D^2P^2}\right)^2,
\end{eqnarray}
with
\begin{equation}
\label{Eq:f}
f(z^{\ast},P)=a(z^{\ast}) \sigma_1(z^{\ast},P)
+ b(z^{\ast})\sigma_2(z^{\ast},P).
\end{equation}
Furthermore, $\delta \sigma_{\rm LR}^{\rm pol}$ 
and $\delta \sigma_{\rm RL}^{\rm pol}$ are obtained from
$\delta \sigma_{\rm LL}^p$ and $\delta \sigma_{\rm RR}^p$,
respectively, by substituting $a(z^{\ast})\leftrightarrow b(-z^{\ast})$.
Numerically, for explicit assessments of the reach on
$\Lambda_{\alpha\beta}$, we shall work out the example of
$\vert P_e\vert=0.9$ with $\delta P_e/ P_e=0.5\%$ 
as currently attainable at the SLC \cite{SLC}, 
and $\vert P_{\bar e}\vert=0.6$ \cite{Accomando}.
This corresponds to the effective polarization 
$P_{\rm eff}=P=0.974$ and $D=1.54$.
Clearly, introducing positron polarization may amount to 
a sort of ``noise'', unless its magnitude is known with some precision.
Since, at present, information on the achievable precision on the positron 
polarization is unknown, in our numerical analysis we shall vary 
$\delta P_{\bar e}/P_{\bar e}$ in a range up to a few tens of percent.

We start by considering, as a first example, electrons that are polarized,
but unpolarized positrons, ($|P_e|,|P_{\bar e}|$) = ($0.9, 0.0$), and
then we discuss the case of both initial beams polarized with 
($|P_e|,|P_{\bar e}|$) = ($0.9, 0.6$). 
Also, as anticipated, we assume half the total integrated luminosity 
quoted above for each value of the effective polarization, 
$P_{\rm eff}=\pm P$.
We focus on the impact of finite polarization uncertainties 
on the sensitivity of the helicity cross sections $\sigma_{\alpha\beta}$ 
to the contact interaction (\ref{lagra}) 
and the corresponding reach on the mass scales $\Lambda_{\alpha\beta}$ that,
as discussed in the previous section, is determined by the uncertainties
$\delta\sigma_{\alpha\beta}$ {\it via} Eqs.~(\ref{signif}), (\ref{deltasig}) 
and (\ref{poldelta}).

\begin{figure}[htb]
\refstepcounter{figure}
\label{Fig:sens-rat-neg}
\addtocounter{figure}{-1}
\begin{center}
\setlength{\unitlength}{1cm}
\begin{picture}(12,6)
\put(-2.4,0.0)
{\mbox{\epsfysize=6.4cm\epsffile{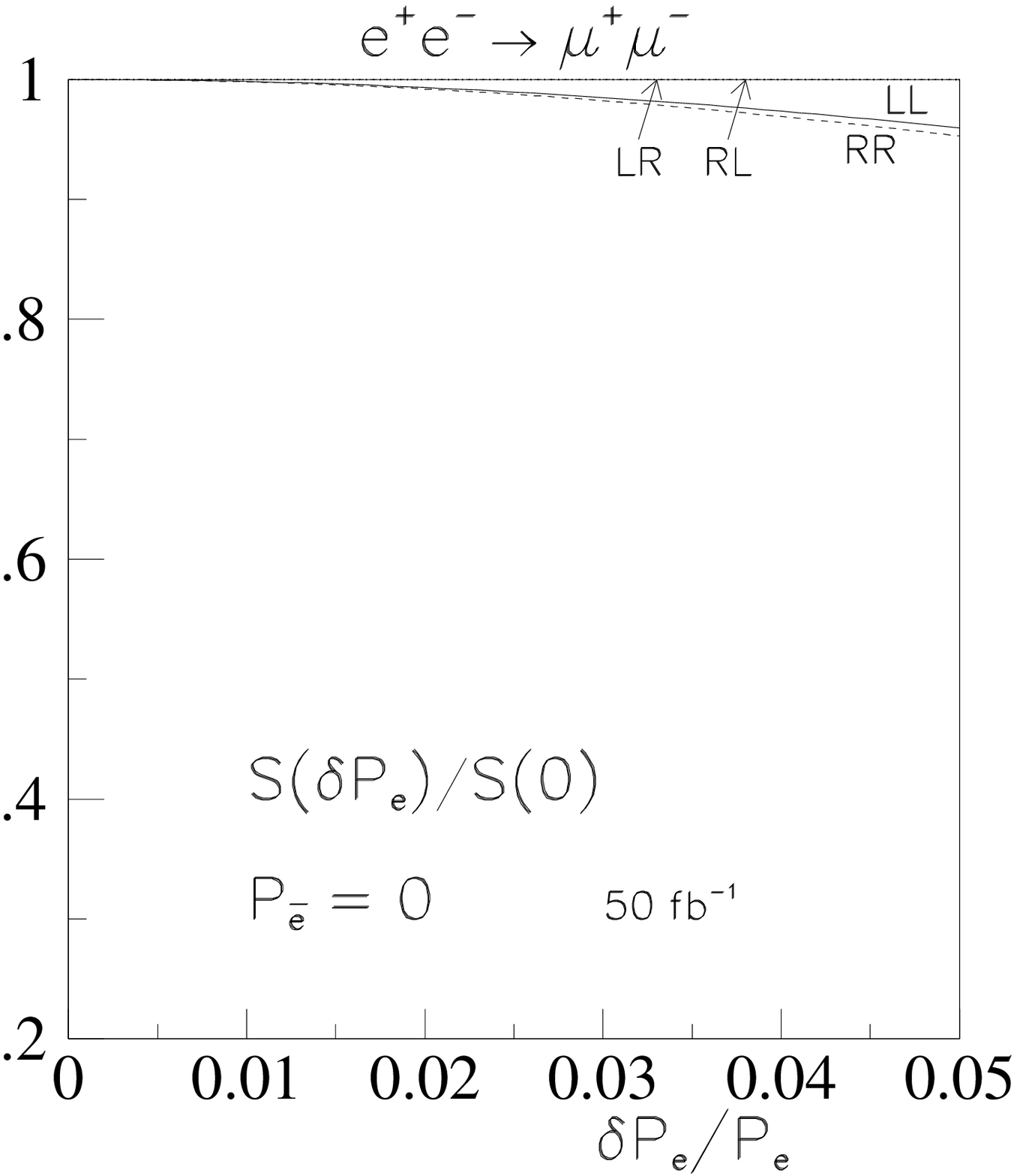}}
 \mbox{\epsfysize=6.4cm\epsffile{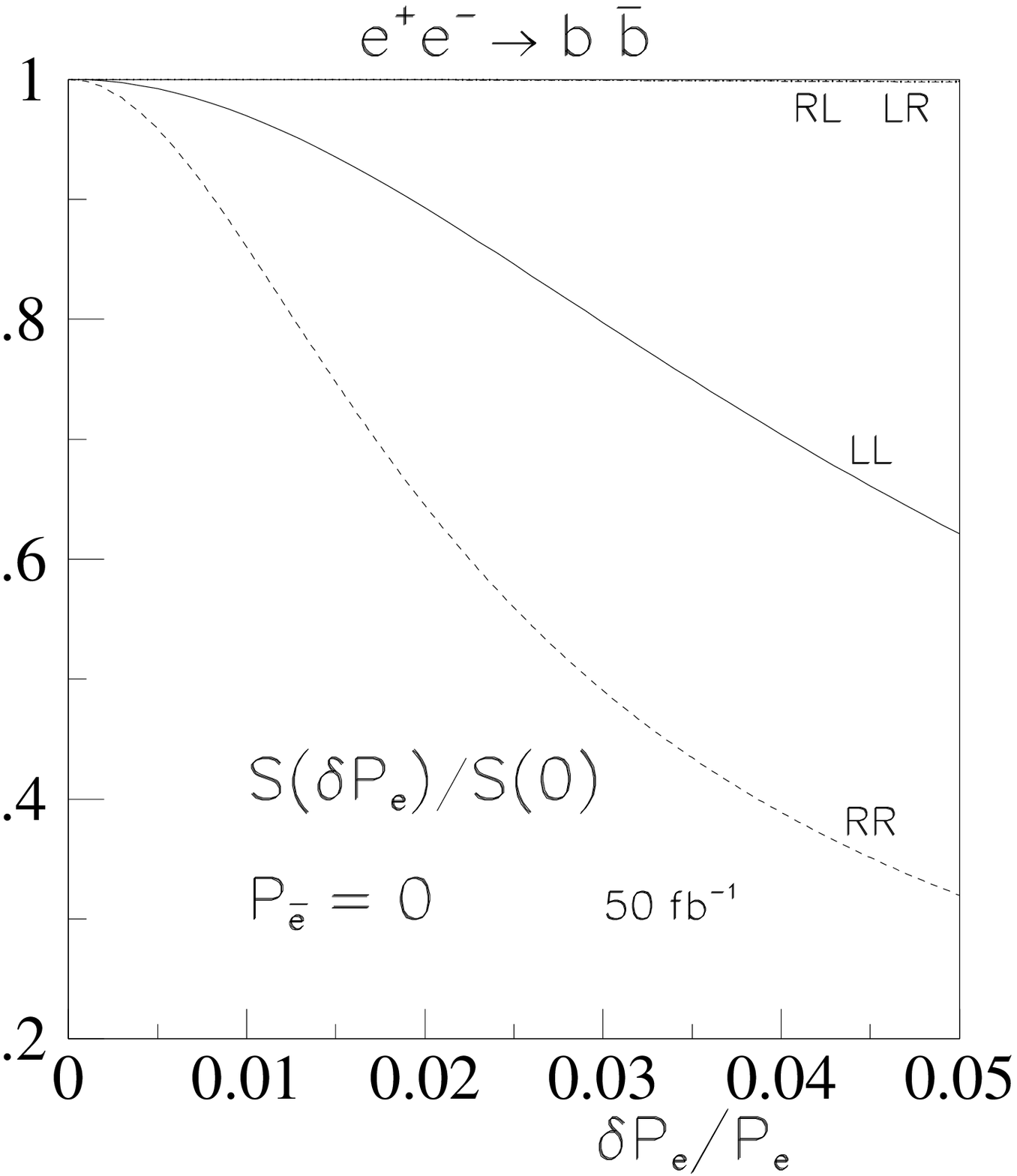}}
 \mbox{\epsfysize=6.4cm\epsffile{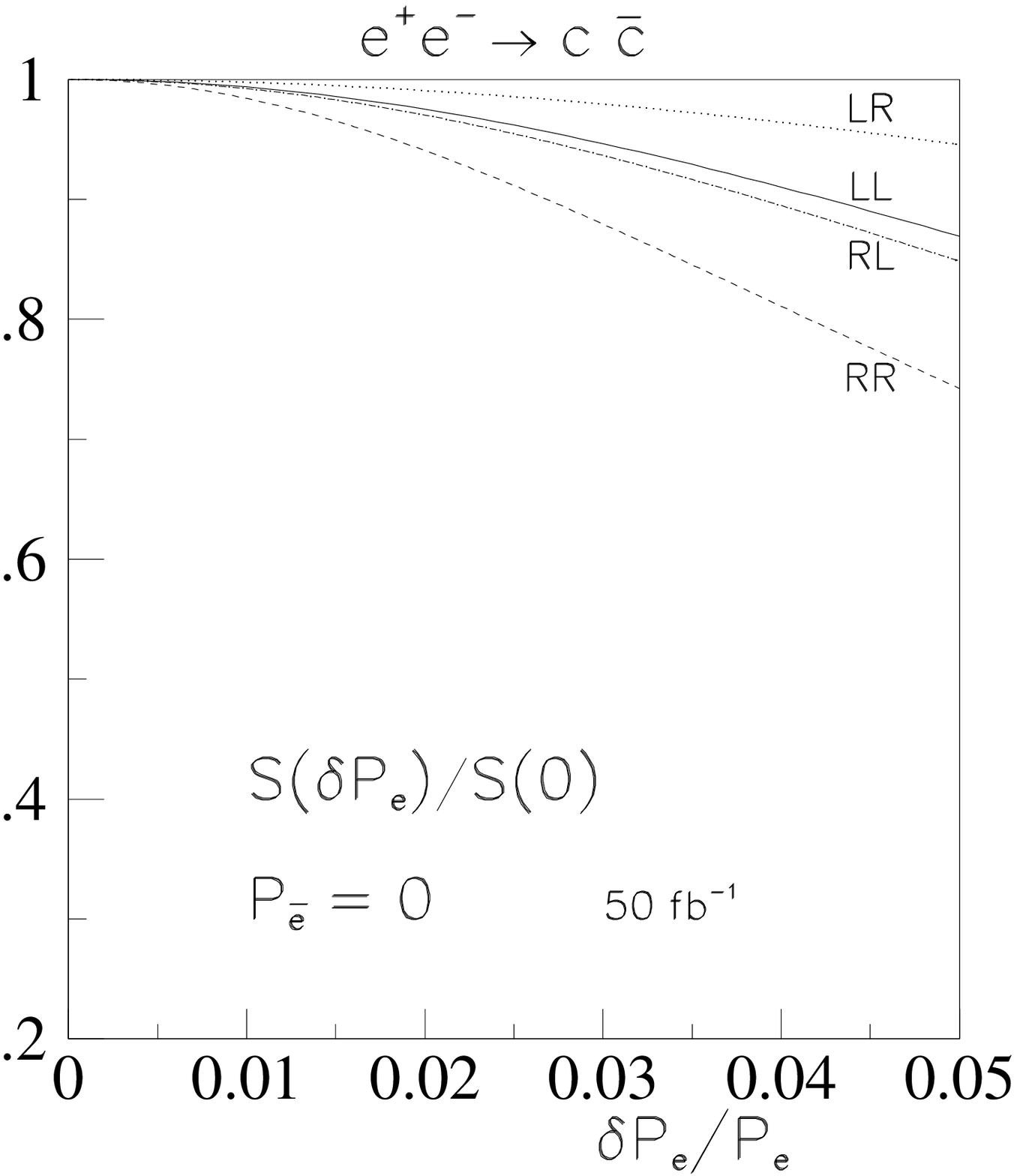}}}
\end{picture}
\vspace*{-3mm}
\caption{
Ratios of sensitivities of helicity cross sections to contact interaction 
parameters as a function of $\delta P_e/P_e$
compared to no electron polarization uncertainty,
for $|P_e|=0.9$, $|P_{\bar{e}}|=0.0$
and ${\cal L}_{\rm int}=50\ {\rm fb}^{-1}$. Helicity configurations are 
indicated.}
\end{center}
\end{figure}

In the starting example, with polarized electrons and unpolarized 
positrons, we compare the relative 
deviations $\delta\sigma_{\alpha\beta}/\sigma_{\alpha\beta}
\simeq \delta\sigma^{\rm SM}_{\alpha\beta}/\sigma^{\rm SM}_{\alpha\beta}$
for finite $\delta P_e$ with the case of the same $P_e$, 
but $\delta P_e=0$, studied in \cite{BOPP-99}. 
The ratio of the sensitivity (\ref{signif}) in the two cases,
determining the effect of the electron polarization uncertainty
introduced via Eq.~(\ref{poldelta}), is shown in Fig.~\ref{Fig:sens-rat-neg},
for ${\cal L}_{\rm int}=50\ {\rm fb}^{-1}$.
This figure is obtained using the optimization procedure, 
and the determination of the relevant $z^*_{\rm opt}$, 
outlined in the previous section.
The sensitivity, via its square root, determines the reach in 
$\Lambda_{\alpha\beta}$.
For the $\mu^+\mu^-$ final state, and
LL and RR helicity configurations, the effect of $\delta P_e$
determining $\delta\sigma_{\alpha\beta}^{\rm pol}$ in (\ref{poldelta})
is found to change $\delta\sigma_{\rm LL}$ and $\delta\sigma_{\rm RR}$
as given by (\ref{uncet1})--(\ref{delsi1}) and the stated input values
by a really modest amount, of the order of a fraction of a \%, 
unless $\delta P_e/P_e$ exceeds 3--4\%,
whereas for $\delta\sigma_{\rm LR}$ and 
$\delta\sigma_{\rm RL}$ there is no change at all. 
The reason for this can be found in Eqs.~(\ref{Eq:uncert-P}) and
(\ref{Eq:f}).
Indeed, within the set of assumptions leading to those equations,
one has numerically:
\begin{eqnarray}
\left(\delta \sigma_{\rm LL, RR}^{\rm pol}\right)^2
&\sim&[(\sigma_{\rm LL}-\sigma_{\rm RR})
\mp P_{\bar e} P(\sigma_{\rm LL}+\sigma_{\rm RR})]^2(\delta P_e)^2 
\nonumber \\
&+&[(\sigma_{\rm LL}-\sigma_{\rm RR})
\pm P_e P(\sigma_{\rm LL}+\sigma_{\rm RR})]^2(\delta P_{\bar e})^2, 
\label{Eq:uncert-P1-P2}
\end{eqnarray}
independent of $z^*$,
and similar expressions for $\delta \sigma_{\rm LR, RL}^{\rm pol}$
with the substitutions $\text{LL,RR}\to\text{LR,RL}$.
Thus, for $P_{\bar e}=\delta P_{\bar e}=0$,
$\delta \sigma_{\rm LL}^{\rm pol}
\sim\delta \sigma_{\rm RR}^{\rm pol}
\sim\sigma_{\rm LL}^{\rm SM}-\sigma_{\rm RR}^{\rm SM}$,
which for final-state muons vanishes in the limit of $\sin^2\thW\to0.25$, 
whereas $\delta \sigma_{\rm LR}^{\rm pol}
\sim\delta \sigma_{\rm RL}^{\rm pol}
\sim\sigma_{\rm LR}^{\rm SM}-\sigma_{\rm RL}^{\rm SM}=0$.
It should be stressed that this lack of sensitivity to $\delta P_e$
depends on having no positron polarization, $P_{\bar{e}}=0$.
For quarks, the corresponding differences of helicity cross sections
do not vanish, and the effect of $\delta P_e$ is to yield a
non-zero $\delta \sigma_{\alpha\beta}^{\rm pol}$. 
The contribution of $\delta P_e$ to the helicity cross section uncertainty,
$\delta \sigma_{\alpha\beta}^{\rm pol}$, is still quite small with
respect to the total uncertainty, as long as $\delta P_e/P_e$
remains less than 2--3\%, except for the LL and RR cases of
$b \bar b$ final states. 
\begin{figure}[htb]
\refstepcounter{figure}
\label{Fig:sens-rat-neg-500}
\addtocounter{figure}{-1}
\begin{center}
\setlength{\unitlength}{1cm}
\begin{picture}(12,6)
\put(-2.4,0.0)
{\mbox{\epsfysize=6.4cm\epsffile{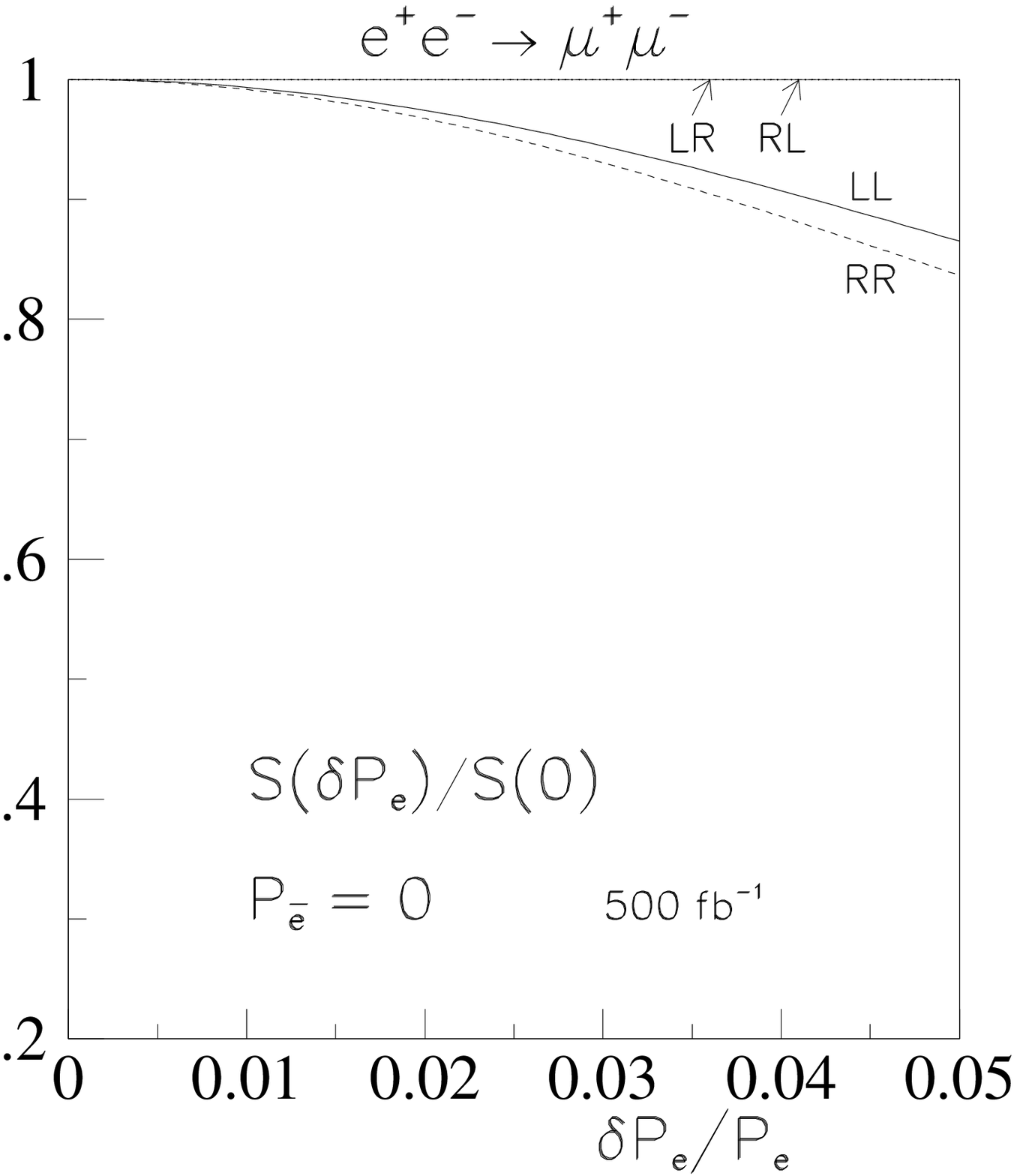}}
 \mbox{\epsfysize=6.4cm\epsffile{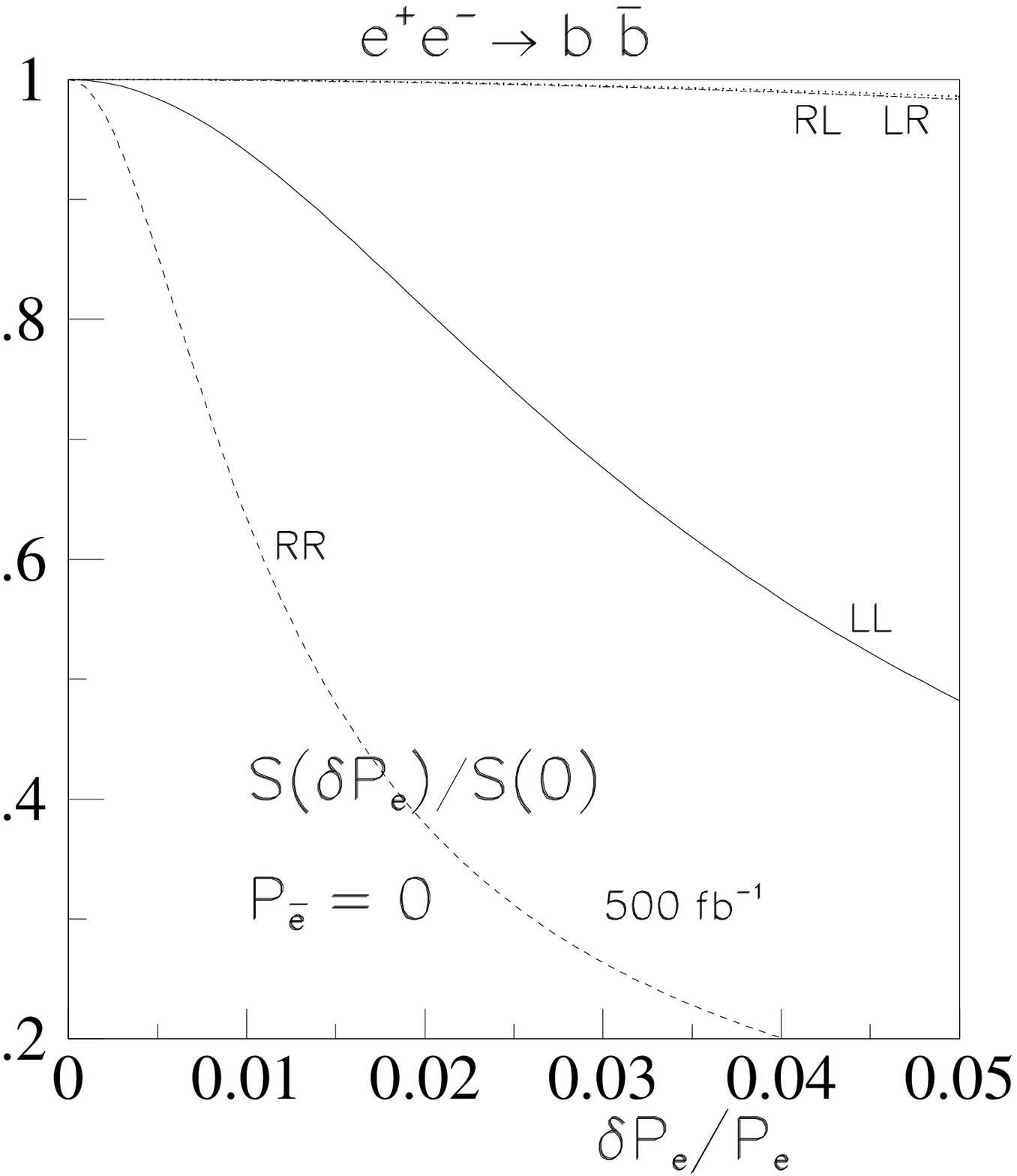}}
 \mbox{\epsfysize=6.4cm\epsffile{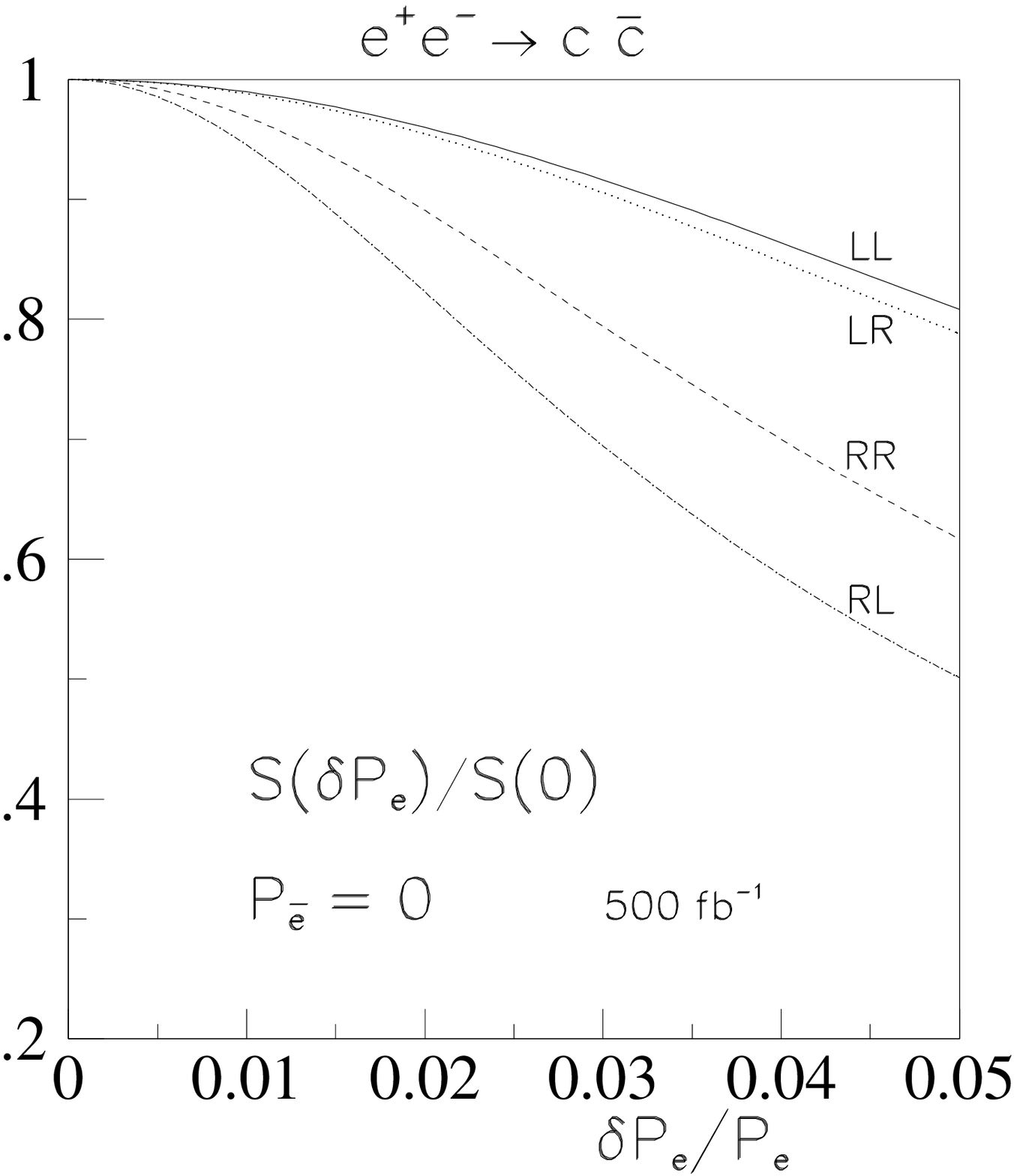}}}
\end{picture}
\vspace*{-3mm}
\caption{
Same as Fig.~\ref{Fig:sens-rat-neg}, 
but for ${\cal L}_{\rm int}=500\ {\rm fb}^{-1}$.}
\end{center}
\end{figure}
For higher luminosity, the curves become steeper, i.e., the sensitivity
deteriorates faster with loss of polarization accuracy, 
see Fig.~\ref{Fig:sens-rat-neg-500}.

\begin{figure}[htb]
\refstepcounter{figure}
\label{Fig:del-P-pos-50}
\addtocounter{figure}{-1}
\begin{center}
\setlength{\unitlength}{1cm}
\begin{picture}(12,6)
\put(-2.7,0)
{\mbox{\epsfysize=6.7cm\epsffile{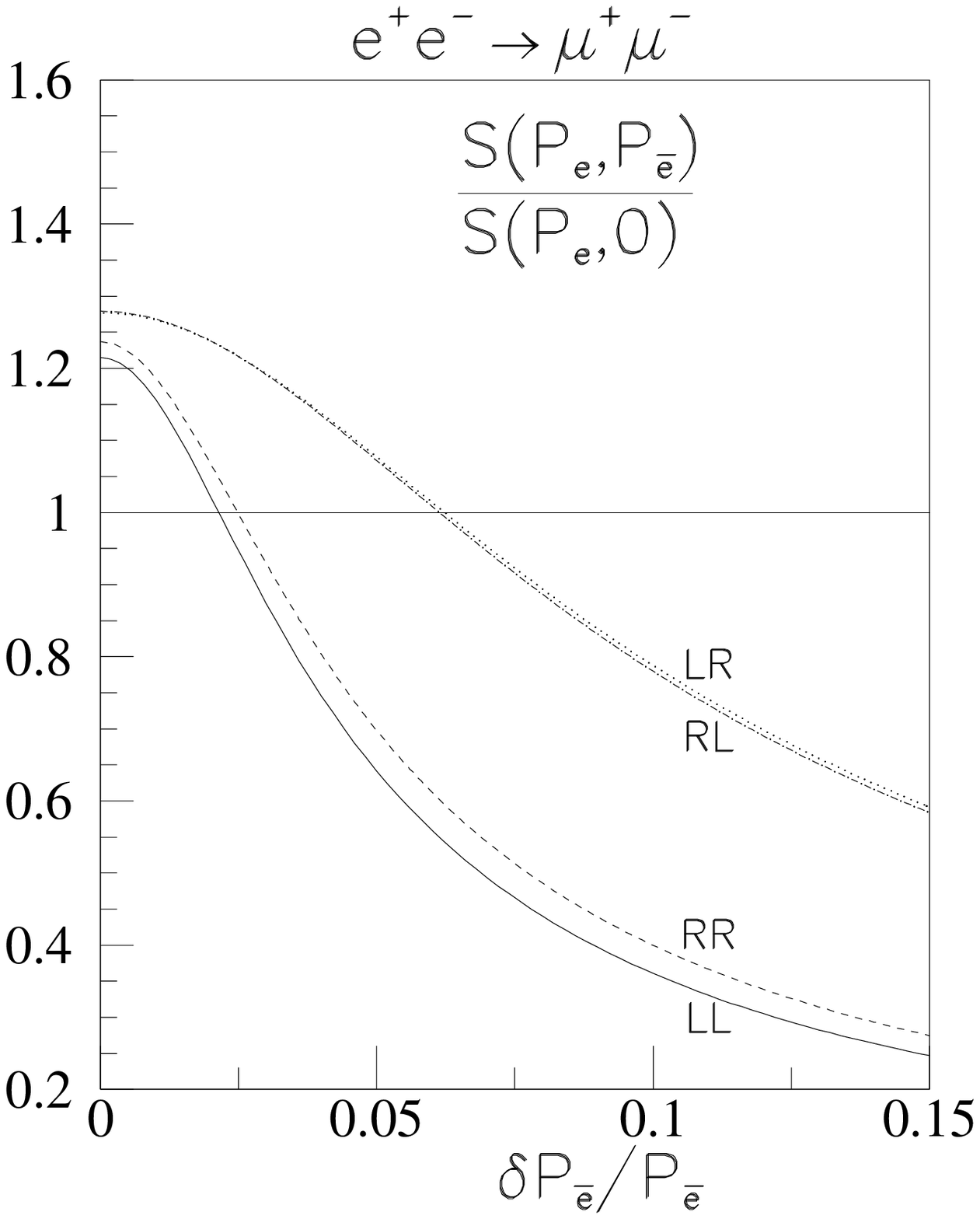}}
 \mbox{\epsfysize=6.7cm\epsffile{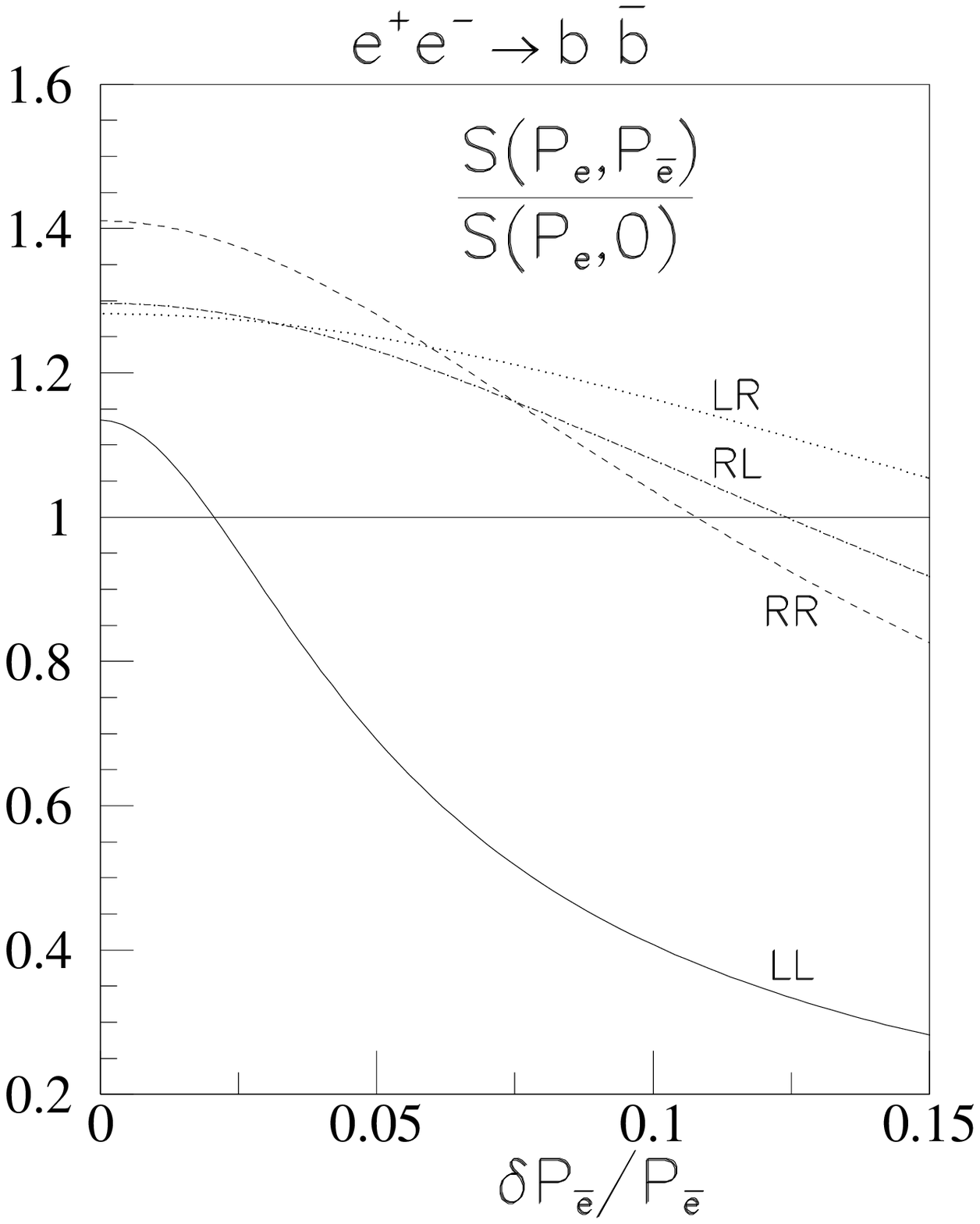}}
 \mbox{\epsfysize=6.7cm\epsffile{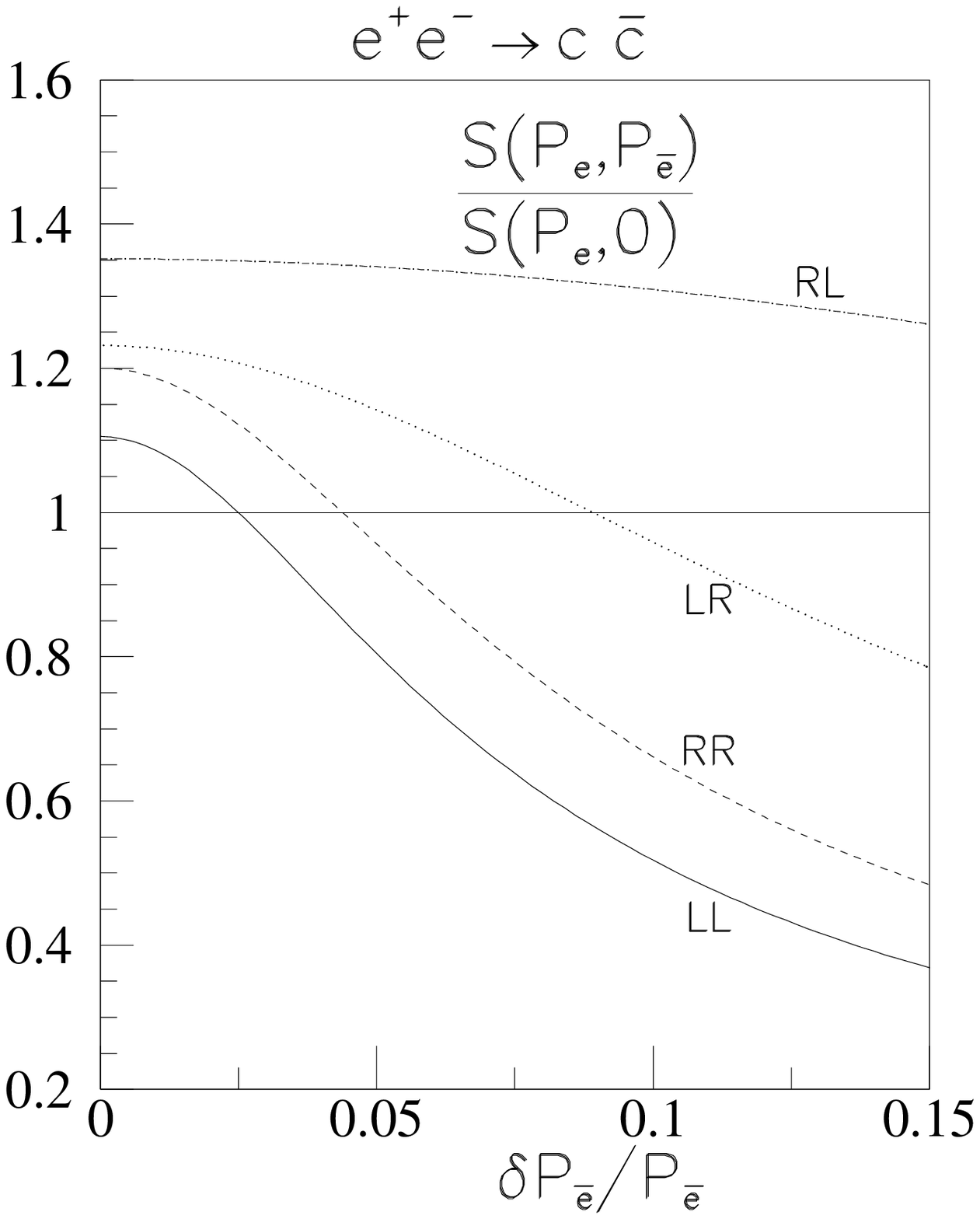}}}
\end{picture}
\vspace*{-3mm}
\caption{
Ratios of sensitivities of helicity cross sections to contact interaction 
parameters as a function of $\delta P_{\bar{e}}/P_{\bar{e}}$
provided by positron polarization compared to no positron polarization,
for $|P_e|=0.9$, $\delta P_e/P_e=0.5\%$ and $|P_{\bar{e}}|=0.6$, 
${\cal L}_{\rm int}=50~{\rm fb}^{-1}$.}
\end{center}
\end{figure}
\begin{figure}[htb]
\refstepcounter{figure}
\label{Fig:del-P-pos-500}
\addtocounter{figure}{-1}
\begin{center}
\setlength{\unitlength}{1cm}
\begin{picture}(12,6)
\put(-2.7,0)
{\mbox{\epsfysize=6.7cm\epsffile{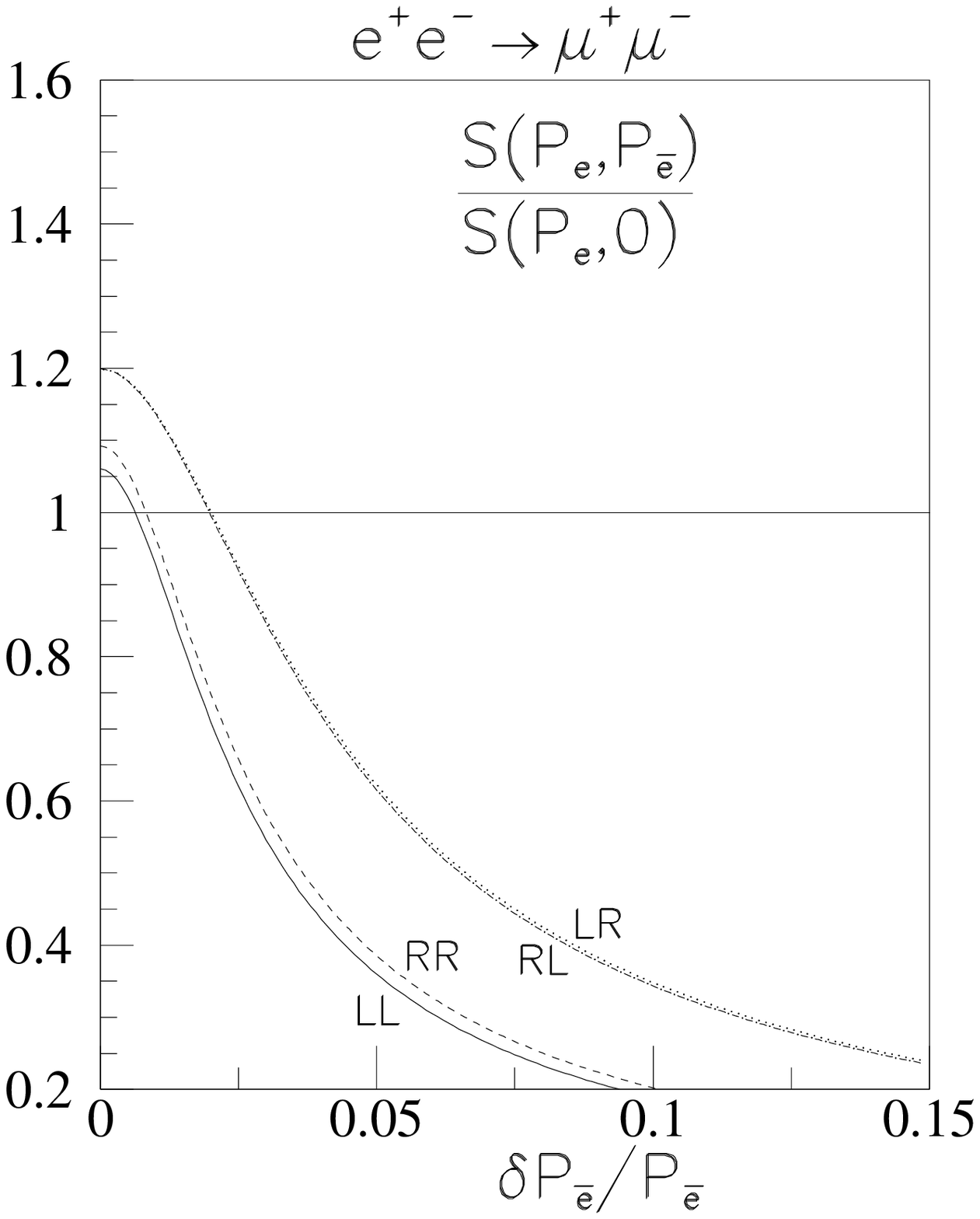}}
 \mbox{\epsfysize=6.7cm\epsffile{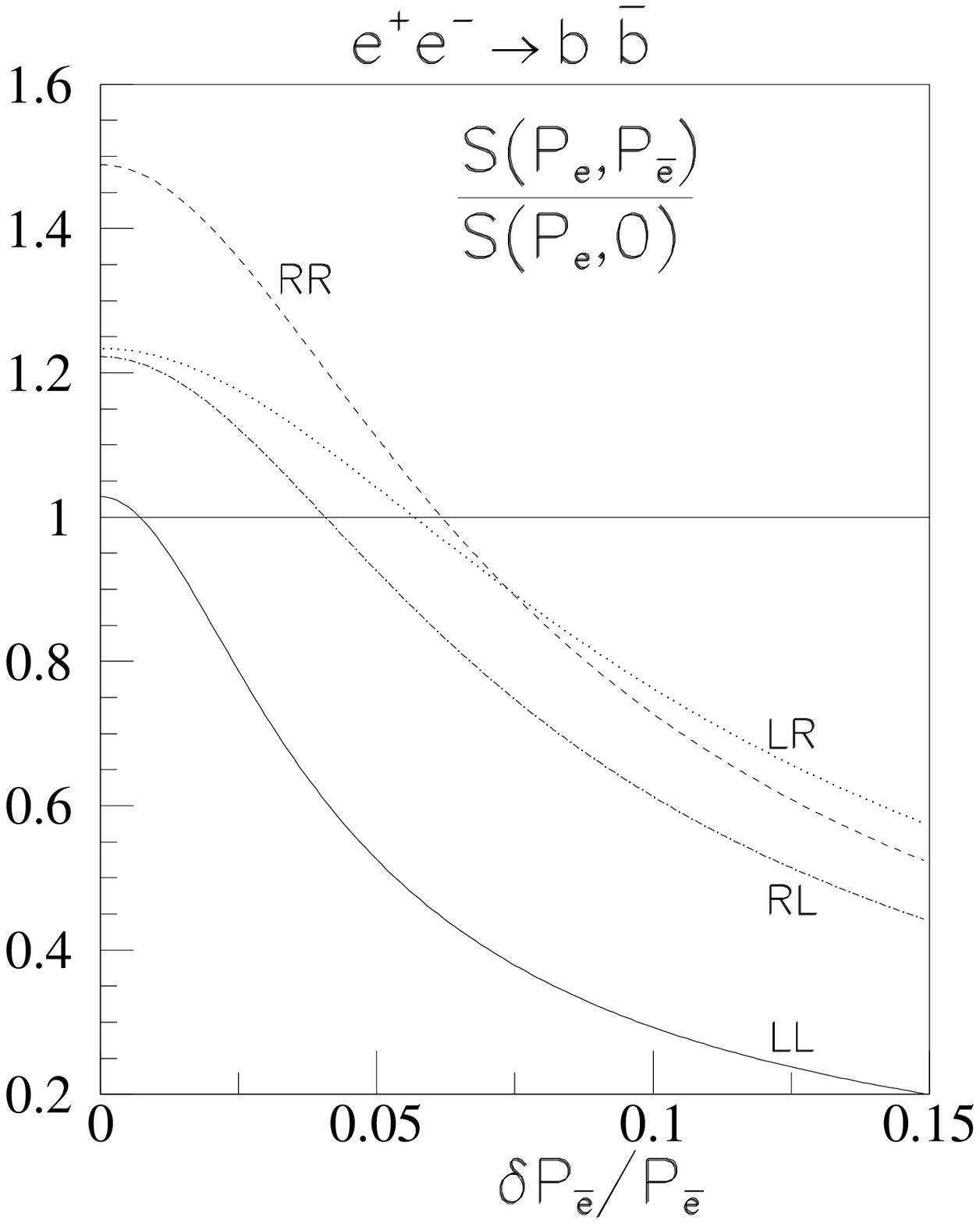}}
 \mbox{\epsfysize=6.7cm\epsffile{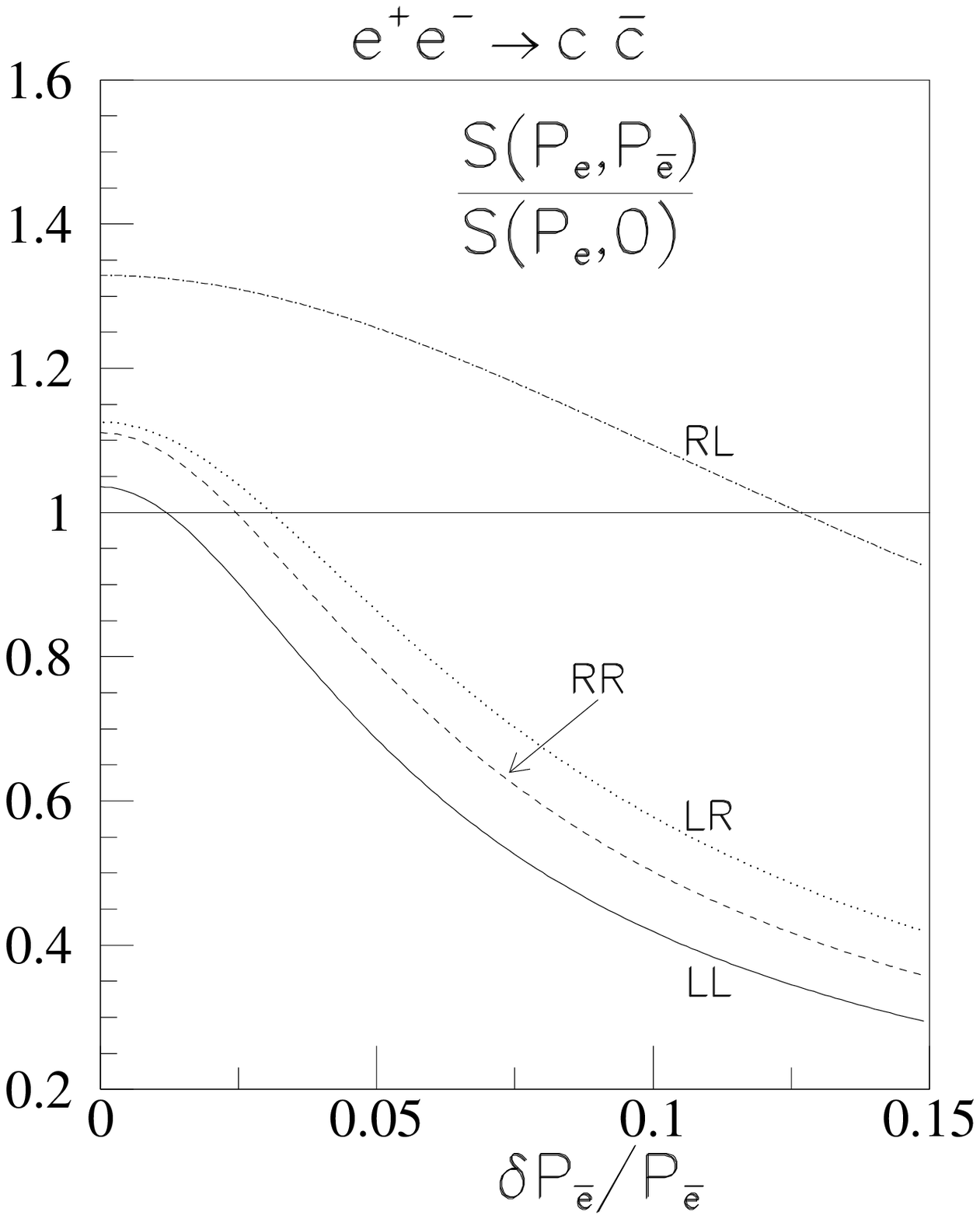}}}
\end{picture}
\vspace*{-3mm}
\caption{
Same as Fig.~\ref{Fig:del-P-pos-50}, for
${\cal L}_{\rm int}=500~{\rm fb}^{-1}$.}
\end{center}
\end{figure}

Turning to the case of both positron and electron longitudinal polarization, 
and referring to Eqs.~(\ref{s+}) and (\ref{s-}), in the chosen helicity
configuration where $P_e P_{\bar{e}}<0$, one has  $D>1$ and 
$|P_{\rm eff}|>\max(|P_e|,|P_{\bar{e}}|)$, 
and in principle one could expect on statistical grounds an increase of the 
sensitivity due to the polarization of positrons, provided
the luminosity remains the same.
However, this improvement from positron polarization is obtained up to
a maximum value of $\delta P_{\bar{e}}/P_{\bar{e}}$, above which
there is no benefit, but, actually, a worsening of the sensitivity.

Indeed, it is instructive to compare the sensitivity of the helicity cross
sections to four-fermion contact interactions for both beams 
polarized with that obtained with just one beam polarized.
This comparison is expressed in terms of ratios of sensitivities 
as a function of the positron polarization uncertainty,
$\delta P_{\bar{e}}/P_{\bar{e}}$, for ${\cal L}_{\rm int}=50~{\rm fb}^{-1}$
in Fig.~\ref{Fig:del-P-pos-50} for lepton and quark final states. 
It is seen that if 
$|\delta P_{\bar{e}}/P_{\bar{e}}|\approx|\delta P_{e}/P_{e}|=0.5\%$, 
the advantage of positron polarization 
manifests itself in an increase in sensitivity by 10--40\% depending on 
the helicity configuration and the final state. 
However, this ratio drops with increasing
$\delta P_{\bar{e}}/P_{\bar{e}}$, and at those positron polarization
uncertainties where it becomes less than unity, the advantage of positron
polarization disappears. This useful region of the precision
$\delta P_{\bar e}/P_{\bar e}$ ranges from 2\% up to beyond 20\% 
depending on the reaction and helicity combination.

This dependence on $\delta P_{\bar e}$ can be qualitatively understood 
from Eq.~(\ref{Eq:uncert-P1-P2}).
In the case of muons (as opposed to quarks), the first term 
(proportional to $(\delta P_e)^2$) is relatively small (since
we consider $P_{\bar e}$ considerably less than $P_e$), 
and the second term involving $(\delta P_{\bar e})^2$
becomes important already at small values of $\delta P_{\bar e}$.
This explains why the curves (see Fig.~\ref{Fig:del-P-pos-50}) are
rather steep. 
Other properties of Fig.~\ref{Fig:del-P-pos-50} 
are also seen to follow from Eq.~(\ref{Eq:uncert-P1-P2}):
(i) for $e^+e^-\to\mu^+\mu^-$, the dependence on $\delta P_{\bar e}$
is the same for the LR and RL cross sections,
as well as for the LL and RR ones;
(ii) for $e^+e^-\to b\bar b$, the dependence on $\delta P_{\bar e}$
is relatively weak for the LR and RL cross sections since these
cross sections are small;
(iii) for $e^+e^-\to c\bar c$, the dependence on $\delta P_{\bar e}$
is much weaker for the RL than for the LR cross section since
$\sigma_{\rm LR}$ is bigger than $\sigma_{\rm RL}$, leading to
a cancellation in one case and not in the other.

At higher luminosity, all the curves become more steep, since the
uncertainty due to the polarization becomes more important w.r.t.\
the statistical uncertainty.
For example, at $\Lumint=500\ {\rm fb}^{-1}$, 
as Fig.~\ref{Fig:del-P-pos-500} shows,
for muon final states the positron polarization 
(at a value $P_{\bar e}=0.6$) stops being useful for the RR and LL
cross sections at $\delta P_{\bar e}/P_{\bar e}=0.8\%$ and 0.6\%,
respectively.

In the next section we are going to conclude our numerical discussion by 
explicitly deriving the reach on the mass scales $\Lambda_{\alpha\beta}$ 
obtainable in the case where the uncertainty on the electron and
positron longitudinal polarizations are, respectively, 0.5\% and
1\%, for the two values $\Lumint=50\, \text{fb}^{-1}$ and 
$\Lumint=500\, \text{fb}^{-1}$.

\section{Bounds on $\Lambda_{\alpha\beta}$}
As a preliminary step in the derivation of the constraints on 
$\Lambda_{\alpha\beta}$, we
show in Fig.~\ref{Fig:del-sig} 
the relative uncertainties
$\delta\sigma_{\alpha\beta}/\sigma_{\alpha\beta}$ as functions of $z^*$, for
the lower option for the luminosity.            
\begin{figure}[thb]
\refstepcounter{figure}
\label{Fig:del-sig}
\addtocounter{figure}{-1}
\begin{center}
\setlength{\unitlength}{1cm}
\begin{picture}(12.0,6.5)
\put(-2.5,0.0){
\mbox{\epsfysize=6.4cm\epsffile{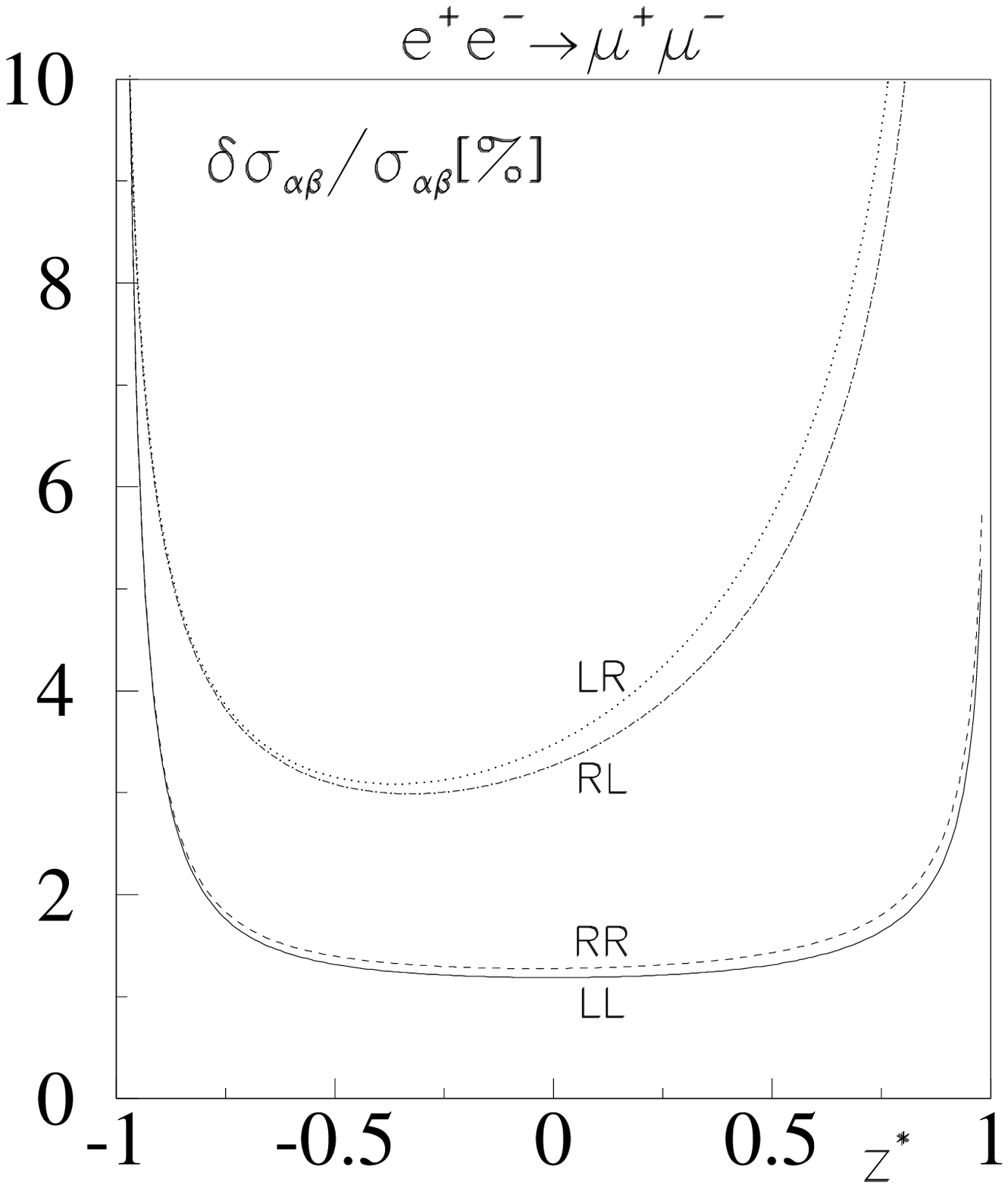}}
\mbox{\epsfysize=6.4cm\epsffile{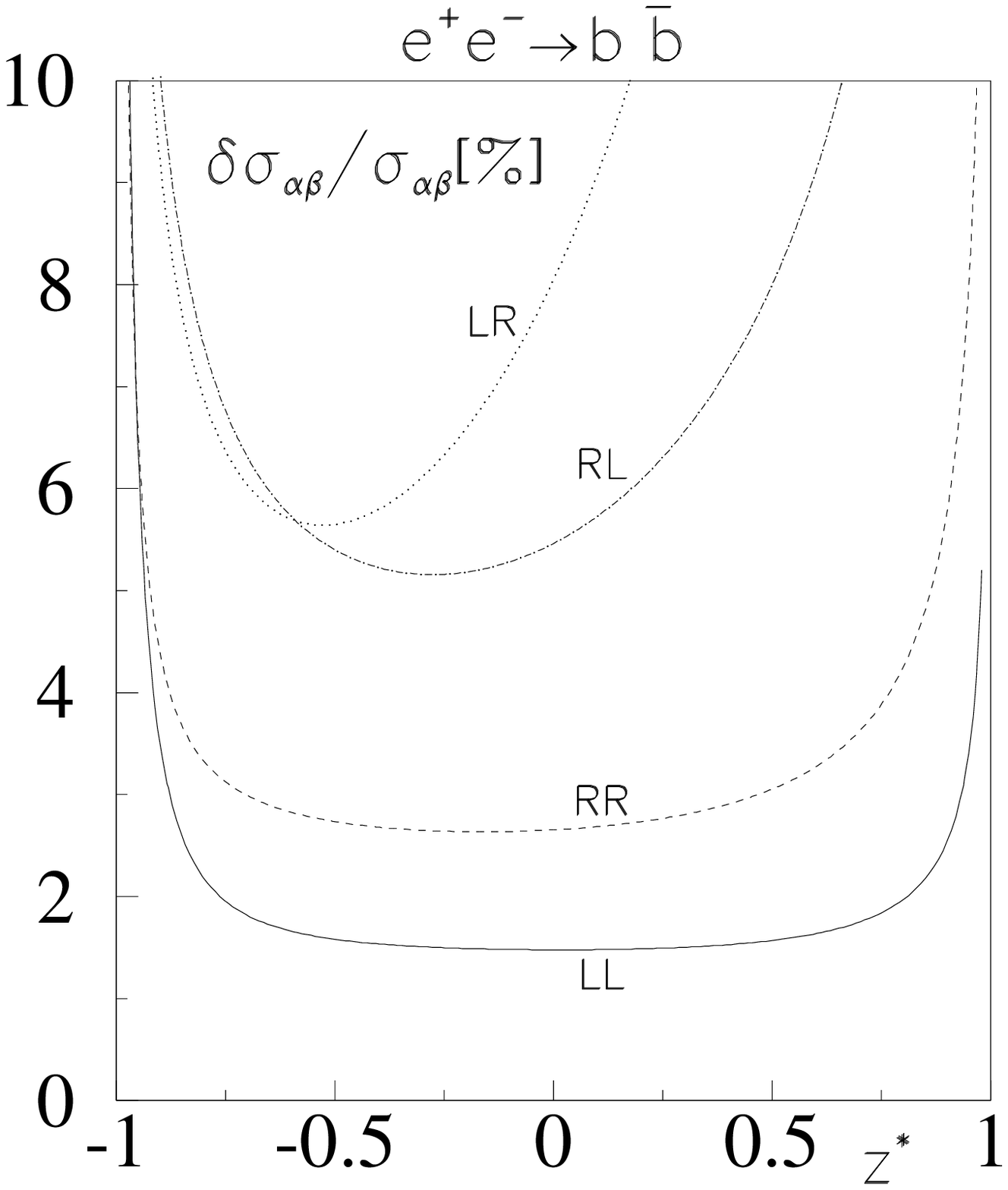}}
\mbox{\epsfysize=6.4cm\epsffile{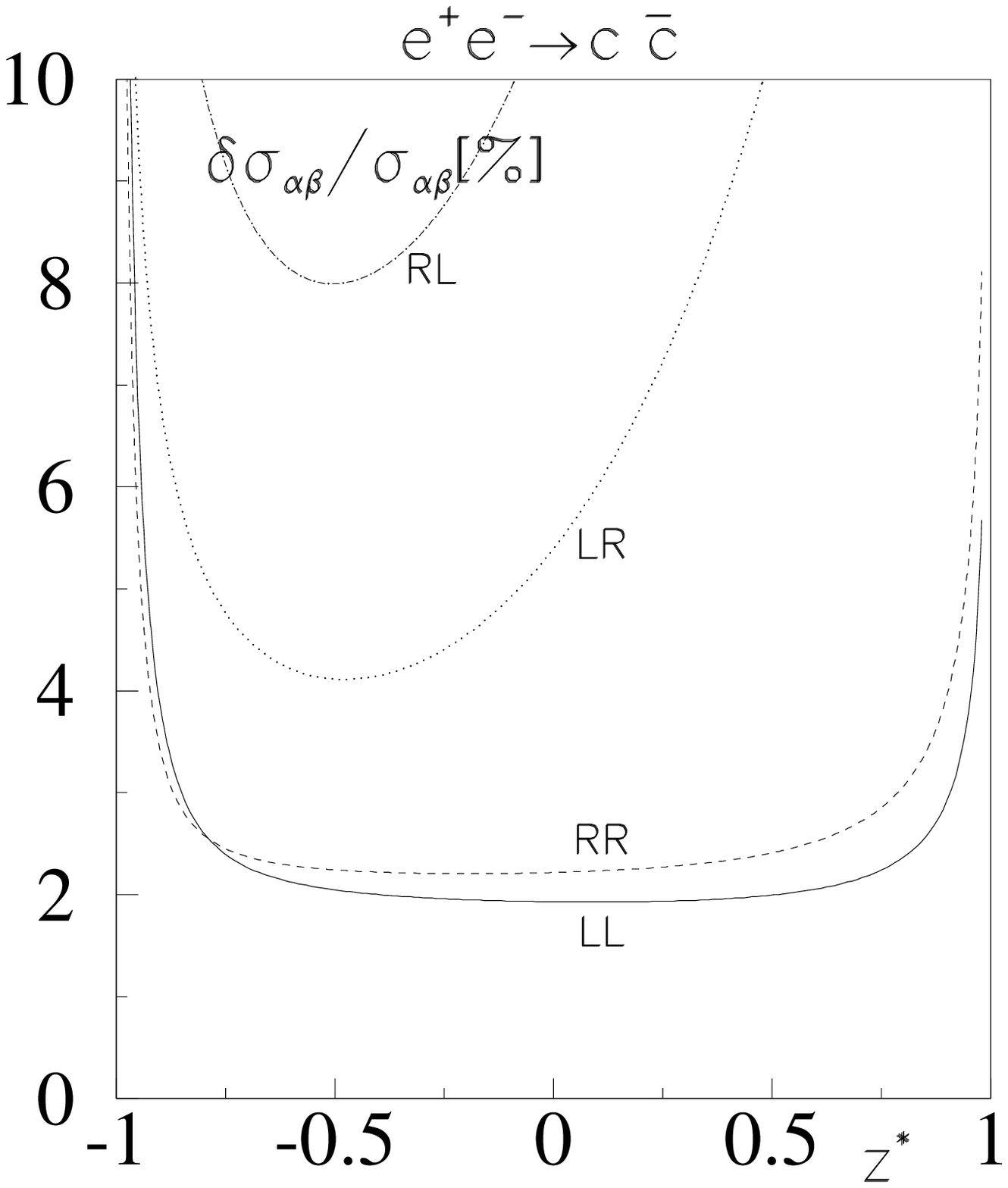}}}
\end{picture}
\vspace*{-2mm}
\caption{
The uncertainty on the helicity cross sections $\sigma_{\alpha\beta}$
in the SM
as a function of $z^*$ for the process $e^+e^-\to\mu^+\mu^-$ at
$P_e=0.9$, $P_{\bar e}=0.6$ and
$\sqrt{s}=0.5$~TeV, $\Lumint=50\ \mbox{fb}^{-1}$.}
\end{center}
\end{figure}
The optimal values of $z^*$ where the sensitivity is maximum can be easily 
read off from these figures, and in Table~1 we report such $z^*_{\rm opt}$ 
for the two different values of the luminosity.  

Numerical constraints on the four-fermion contact interactions of 
Eq.~(\ref{lagra}) are obtained from a $\chi^2$ analysis of data on 
each helicity cross section, with (see Eq.~(\ref{signif})):
\begin{equation}
\label{chisq}
\chi^2
=\left(\frac{\Delta\sigma_{\alpha\beta}}{\delta\sigma_{\alpha\beta}}\right)^2.
\end{equation}
Bounds on the allowed values of the contact interaction
parameters from the non-observation of the corresponding deviations within
the expected uncertainty $\delta\sigma_{\alpha\beta}$ are derived by imposing
$\chi^2<\chi^2_{\rm CL}$, where the actual value of $\chi^2_{\rm CL}$
specifies the desired `confidence' level. As Eq.~(\ref{deltasig}) shows,  
the deviations $\Delta\sigma_{\alpha\beta}$ depend on a single `effective' 
contact-interaction free parameter, and therefore 
in such a $\chi^2$ analysis we take $\chi^2_{\rm CL}=3.84$ for 95\% C.L. 
as consistent with a one-parameter fit. 

\begin{table}[ht]
\centering
\caption{Optimal kinematical cut, $z^{\ast}_{\rm opt}$, 
and resulting contact-interaction reach (in TeV) at 95\% C.L.\ 
at the LC with $E_{\rm c.m.}=0.5$~TeV and 
double beam polarization: $|P_e|=0.9$, $(\delta P_e/P_e)=0.5 \%$,
$|P_{\bar e}|=0.6$, $(\delta P_{\bar e}/P_{\bar e})=1.0 \%$.
($\Lambda_{\alpha\beta}$ values in parentheses refer to no optimization,
$z^{\ast}=0$.)}
\medskip
{\small
\begin{tabular}{|c|c|c|c|c|c|}
\hline
&&&&& \\
process & $\Lumint$ & 
$z^\ast, \, \Lambda_{\rm LL}$ & $z^\ast, \, \Lambda_{\rm RR}$ &
$z^\ast, \, \Lambda_{\rm LR}$ & $z^\ast, \, \Lambda_{\rm RL}$ \\
&$[{\rm fb}^{-1}]$&[TeV]&[TeV]&[TeV]&[TeV] \\
\hline
\cline{2-6} 
$\mu^+\mu^-$
 & 50 & 0.00, (43.0) 43.0&-0.04, (43.4) 43.4
      &-0.36, (37.7) 40.0&-0.33, (38.6) 40.4 \\ \cline{2-6} 
 & 500 &0.50, (54.7) 56.9&0.44, (57.0) 58.8
       &-0.49, (57.1) 65.5&-0.46, (58.9) 65.7 \\ \hline
\hline 
$b \bar b$ 
 &50  &0.01, (44.8) 44.8&-0.16, (51.0) 51.2
      &-0.52, (32.9) 39.8&-0.28, (48.2) 49.6 \\ \cline{2-6}
 &500 &0.57, (50.6) 51.8&0.18, (70.5) 70.7
      &-0.63, (44.5) 65.0&-0.42, (72.8) 78.7 \\ \hline
\hline 
$c \bar c$ 
 &50  &0.09, (35.6) 35.6&-0.19, (40.1) 40.2
      &-0.47, (33.6) 38.4&-0.50, (32.0) 37.0 \\ \cline{2-6}
 &500 &0.60, (38.3) 39.0&0.42, (45.7) 46.0 
      &-0.66, (40.1) 53.5&-0.62, (41.9) 57.9 \\ \hline
\end{tabular} } 
\end{table}
The results for the bounds on $\Lambda_{\alpha\beta}$ are reported  
in Table~1.
The table shows that the helicity cross sections $\sigma_{\alpha\beta}$ are 
quite sensitive to contact interactions, with discovery limits that, at the 
highest considered luminosity 500~$\mbox{fb}^{-1}$, can range from 75 up to
150 times the c.m.\ energy, depending on the considered final fermion state.
Indeed, the best sensitivity is achieved for the $\mu^+\mu^-$ and $b\bar{b}$
final states, while the worst one corresponds to the $c\bar{c}$ channel.
A direct comparison with the sensitivity achieved using `conventional'
observables, Eqs.~(\ref{canon})--(\ref{afbpol})\footnote{See, for example,
the results obtained in \cite{riemann} in the context of specific
contact-interaction models.},
is quite difficult and might be unclear, because it depends 
on the assumed model of new physics involved and the kind of parameterization
adopted for the uncertainty.
In this regard, as repeatedly stressed, we point out that the separation
of the helicity cross sections performed here (and the corresponding values 
in Table~1) has the qualitative advantage of providing, by definition, 
unambiguous and model-independent information on the non-standard parameters 
of Eq.~(\ref{lagra}).

\begin{figure}[thb]
\refstepcounter{figure}
\label{Fig:Lambda-mm}
\addtocounter{figure}{-1}
\begin{center}
\setlength{\unitlength}{1cm}
\begin{picture}(8.5,7.8)
\put(-1.0,-1.5){
\mbox{\epsfysize=10.0cm\epsffile{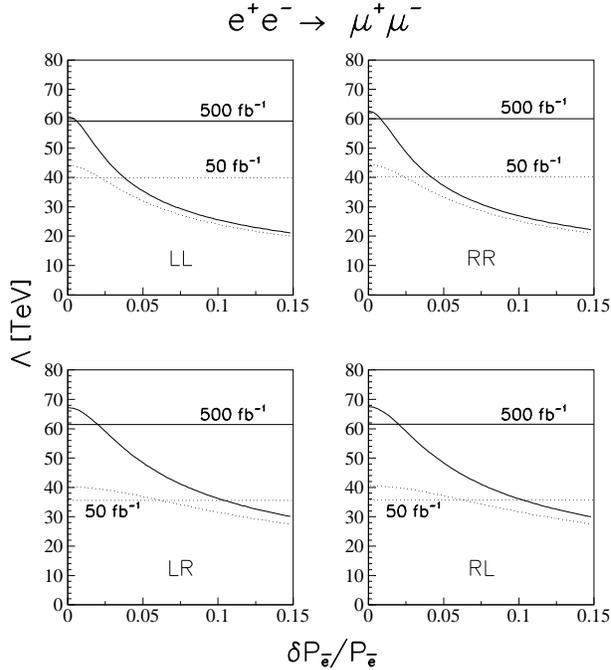}}}
\end{picture}
\vspace*{2mm}
\caption{
Reach in $\Lambda_{\alpha\beta}$ {\it vs.}\ uncertainty in positron 
polarization,
$\delta P_{\bar e}/P_{\bar e}$ for $\mu^+\mu^-$ final states.
Dashed: 50~$\mbox{fb}^{-1}$, solid: 500~$\mbox{fb}^{-1}$.
Horizontal lines: no positron polarization.}
\end{center}
\end{figure}
\begin{figure}[thb]
\refstepcounter{figure}
\label{Fig:Lambda-bc}
\addtocounter{figure}{-1}
\begin{center}
\setlength{\unitlength}{1cm}
\begin{picture}(8.5,7.8)
\put(-4.8,-1.5){
\mbox{\epsfysize=10.0cm\epsffile{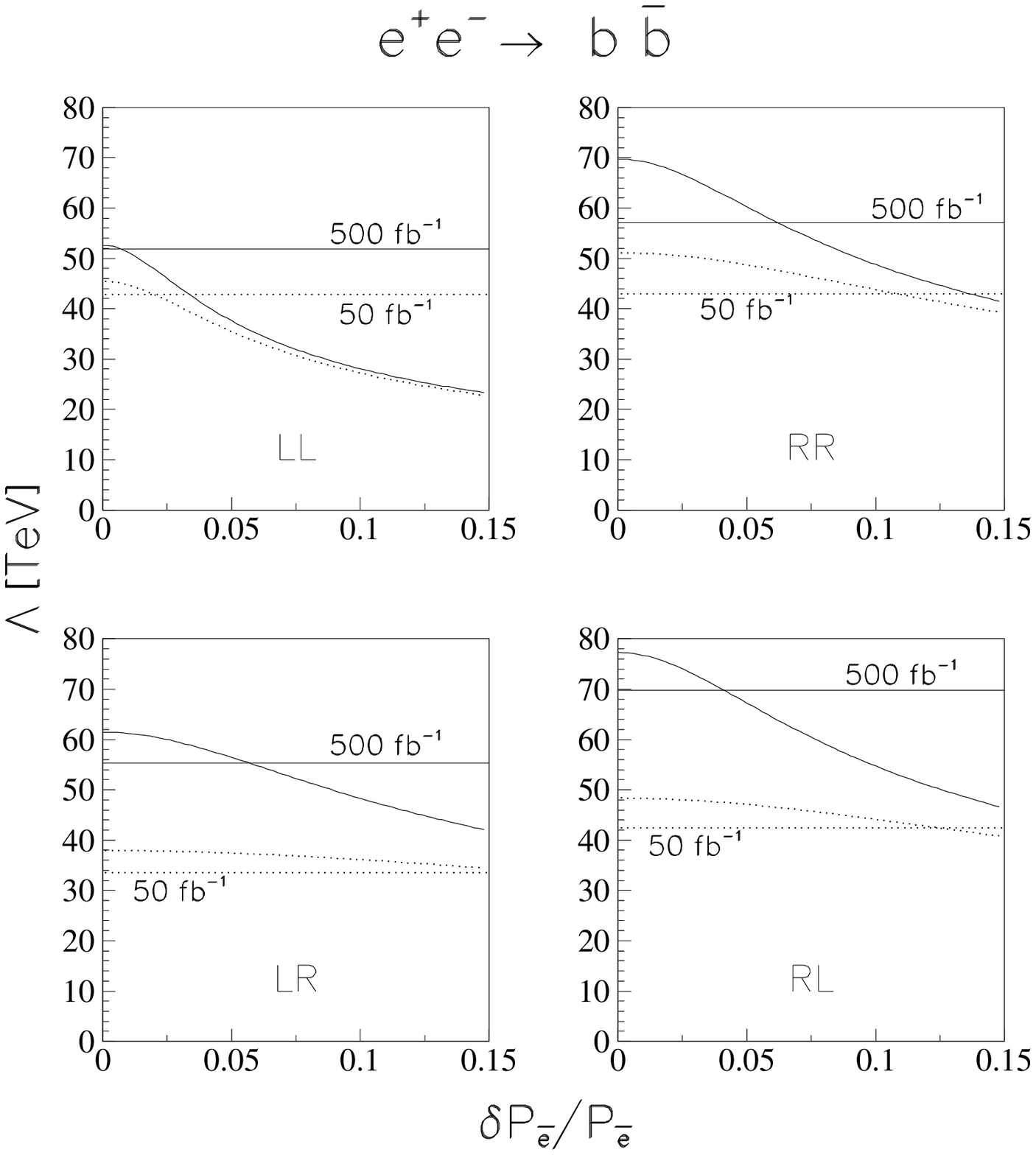}}
\mbox{\epsfysize=10.0cm\epsffile{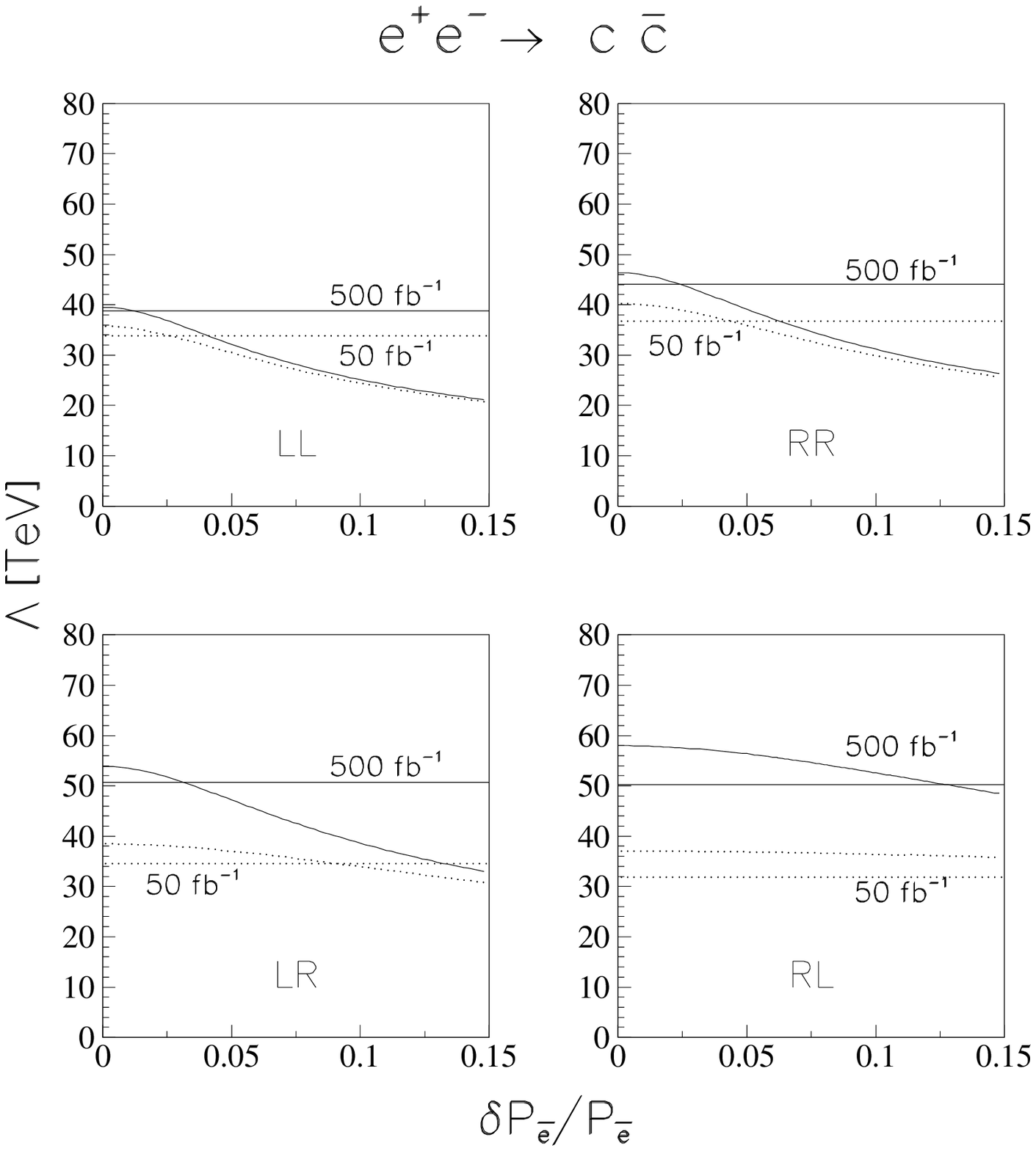}}}
\end{picture}
\vspace*{2mm}
\caption{
Same as Fig.~\ref{Fig:Lambda-mm}, but for $b \bar b$ and $c \bar c$ 
final states.}
\end{center}
\end{figure}


For a sort of contact to the conventional observables 
(\ref{canon})--(\ref{afbpol}), we have reported in Table~1 also 
the limits on $\Lambda_{\alpha\beta}$ obtainable at $z^*=0$ instead of
$z^*=z^*_{\rm opt}$. The results show that, at $z^*=0$, the sensitivity
to $\Lambda_{\rm LR}$ and $\Lambda_{\rm RL}$ would be considerably smaller.
As one can see from the table, the `optimal' choice $z^*=z^*_{\rm opt}$
allows to substantially increase the bounds for the LR and RL cases,
to the level of the LL and RR ones, for which the improvement
is really modest.
This relates to the $z^*$ behavior of the relative uncertainties
on $\sigma_{\alpha\beta}$, that, as is seen in Fig.~\ref{Fig:del-sig}, 
is flat in the latter case and varies
more rapidly around $z^*_{\rm opt}$ in the former one.

As discussed previously, and illustrated in Fig.~\ref{Fig:del-P-pos-50},
the benefit of positron polarization depends on it being known 
with some precision.
We show in Figs.~\ref{Fig:Lambda-mm} and \ref{Fig:Lambda-bc} the effect
of the positron polarization uncertainty on the reach in 
$\Lambda_{\alpha\beta}$, for the two luminosities considered.

We see from these figures that if the positron polarization is known
with high precision, an amount $P_{\bar e}=0.6$ can increase the reach 
in $\Lambda$ by typically 5--25\%. 
The critical level of precision, by which the positron polarization
should be known, in order to be beneficial for contact-interaction
searches, depends very much on the channel considered, as well as
the luminosity. At low luminosities, less polarization precision 
is required for the positron polarization to be useful.

While one polarized beam is a {\it necessity} in order to be able to 
extract the helicity cross sections, the benefit of both beams being 
polarized is less clear. 
For some combinations of final state and helicity channels, 
the increased reach in $\Lambda$ can be considerable, although
half luminosity (and correspondingly reduced number of events) has 
been assumed for the two configurations of electron and positron
beam polarizations.
However, due to the limiting effect of the polarization uncertainties
on the sensitivity (\ref{signif}), such improvements do not seem dramatic.
More luminosity might easily lead to the same gain, especially if the 
positron polarization is only known with a moderate accuracy.

Actually, for full completeness in this regard,
the dependence on the actual value of the uncertainty $\delta^{\rm sys}$ 
in (\ref{delsi1}) should be considered simultaneously with that from 
$\delta\sigma_{\alpha\beta}^{\rm pol}$, as suggested by the combination 
in Eq.~(\ref{poldelta}). Clearly, we should expect reduction of the 
$\Lambda$ reach for increasing $\delta^{\rm sys}$. As an indication, 
by doubling the values of $\delta^{\rm sys}$ with respect to those
listed below Eq.~(\ref{delsi1}), and adopted for the explicit numerical 
example presented here, at $\Lumint=50\ \mbox{fb}^{-1}$ the typical 
effect amounts to a few percent for the LR and RL cases, 
but can be as large as 20\% for the LL and RR combinations. 
This indicates that the latter helicity cross sections are
much more sensitive to systematic uncertainties than the former ones.

Clearly, although these considerations are numerically drawn from 
a specific example using as inputs some particular, hypothetical, 
values of initial beam polarization and corresponding uncertainties, 
and from assumptions on the values
and properties of the uncertainties $\delta^{\rm sys}$ of 
Eq.~(\ref{delsi1}), such conclusions should hold in general.
For a definite, quantitative statement about the relative roles
of statistical and systematic uncertainties (including $\delta P_e$
and $\delta P_{\bar e}$) in the determination of the accuracy on
$\sigma_{\alpha\beta}$ in a realistic experimental situation,
we must wait for more detailed information on the expected 
experimental errors.




\clearpage

\begin{thebibliography}{99}

\bibitem{Eichten}
E. J. Eichten, K. D. Lane, M. E. Peskin,
Phys.\ Rev.\ Lett.\ {\bf 50} (1983) 811; \\
R. R\"uckl, Phys.\ Lett.\ {\bf B 129} (1983) 363.

\bibitem{barger} V. Barger, K. Cheung, K. Hagiwara, 
D. Zeppenfeld, Phys. Rev. D {\bf 57} (1998) 391; \\
D. Zeppenfeld, K. Cheung, preprint MADPH-98-1081, hep-ph/9810277.

\bibitem{altarelli} 
G. Altarelli, J. Ellis, G. F. Giudice, S. Lola, M. L. Mangano, 
Nucl.\ Phys.\ {\bf B 506} (1997) 3; \\
R. Casalbuoni, S. De Curtis, D. Dominici, R. Gatto, Phys.
Lett. {\bf B 460} (1999) 135; \\
V. Barger, K. Cheung, UCD-2000-8, hep-ph/0002259. 

\bibitem{BOPP-99} A.A. Babich, P. Osland, A.A. Pankov, N. Paver,
Phys.\ Lett.\ {\bf B 476} (2000) 95.

\bibitem{Zeppenfeld2}
B. Schrempp, F. Schrempp, N. Wermes, 
D. Zeppenfeld, Nucl. Phys. {\bf B 296} (1988) 1.

\bibitem{Flottmann-Omori} 
K. Fl\"ottmann, preprint DESY 95-064;
K.Fujii, T.Omori, KEK preprint 95-127.

\bibitem{zfitter}
S. Riemann, FORTRAN program ZEFIT Version 4.2; \\
D. Bardin et al., preprint DESY 99-070, hep-ph/9908433.

\bibitem{Hollik}
M. Consoli, W. Hollik, F. Jegerlehner, {\it in} Z physics at LEP1,
G. Altarelli, R. Kleiss, C. Verzegnassi (Eds.), vol.1, p.7, 1989.

\bibitem{Altarelli2}
G. Altarelli, R. Casalbuoni, D. Dominici, F. Feruglio, R. Gatto,
Nucl.\ Phys.\ {\bf B 342} (1990) 15.

\bibitem{Djouadi}
A. Djouadi, A. Leike, T. Riemann, D. Schaile, C. Verzegnassi,
Z. Phys. {\bf C 56} (1992) 289.

\bibitem{Lambda} See, e.g.:
D. Abbaneo et al., The LEP Collaborations, CERN-EP-2000-016; \\ 
J. Breitenweg et al., ZEUS Collaboration, DESY-99-058;\\ 
C. Adloff et al., H1 Collaboration, DESY-00-027.

\bibitem{Damerell}
C. J. S. Damerell, D.J. Jackson,
in {\it Proceedings of the 1996 DPF/DPB Summer
Study on New Directions for High Energy Physics} (Snowmass 96),
Edited by D.G. Cassel, L. Trindle Gennari, R.H. Siemann
(SLAC, 1997) p. 442.

\bibitem{SLC}
M. Swartz, 19th International Symposium on Lepton and Photon 
Interactions at High-Energies (LP 99), Stanford, CA, 9-14 Aug 1999.

\bibitem{Accomando}
E. Accomando et al. (ECFA/DESY LC Physics Working Group),
Phys.\ Rept.\ {\bf 299} (1998) 1.

\bibitem{riemann}
S. Riemann, presentation at the 
``4th International Workshop on Linear Colliders'',
Sitges, Spain, 1999.


\end{thebibliography}
\end{document}